\documentclass[twocolumn]{aastex63}
\usepackage{natbib}
\usepackage{graphicx}
\usepackage[space]{grffile}
\usepackage{multirow} 
\usepackage{changepage} 
\usepackage{chngcntr}
\usepackage{cleveref} 
\usepackage{afterpage} 
\usepackage{hhline}
\usepackage{subfigure}
\usepackage{rotating}

\shorttitle{The Radiation Environments of Middle-Aged F-Type Stars}
\shortauthors{Cruz Aguirre et al.}

\newcommand{\lasp}{Laboratory for Atmospheric and Space Physics, University of Colorado, 600 UCB, Boulder, CO 80309, USA}
\newcommand{\iowa}{The University of Iowa, Department of Physics \& Astronomy, Van Allen Hall, Iowa City, IA 52242, USA}

\begin{document}
\setlength{\skip\footins}{1.2pc plus 5pt minus 2pt}

\title{The Radiation Environments of Middle-Aged F-Type Stars}
\author[0000-0003-4628-8524]{F. Cruz Aguirre}
\affiliation{\lasp}
\affiliation{\iowa}

\author[0000-0002-1002-3674]{K. France}
\affiliation{\lasp}

\author[0000-0001-7131-7978]{N. Nell}
\affiliation{\lasp}

\author{N. Kruczek}
\affiliation{\lasp}

\author[0000-0002-2129-0292]{B. Fleming}
\affiliation{\lasp}

\author[0000-0001-9504-0520]{P. C. Hinton}
\affiliation{\lasp}

\author{S. Ulrich}
\affiliation{\lasp}

\author{P. R. Behr}
\affiliation{\lasp}

\date{\today}

\begin{abstract}

Far ultraviolet (FUV) emission lines from dwarf stars are important driving sources of photochemistry in planetary atmospheres. Properly interpreting spectral features of planetary atmospheres critically depends on the emission of its host star. While the spectral energy distributions (SEDs) of K- and M-type stars have been extensively characterized by previous observational programs, the full X-ray to infrared SED of F-type stars has not been assembled to support atmospheric modeling. On the second flight of the Suborbital Imaging Spectrograph for Transition-region Irradiance from Nearby Exoplanet host stars (SISTINE-2) rocket-borne spectrograph, we successfully captured the FUV spectrum of Procyon A (F5 IV-V) and made the first simultaneous observation of several emission features across the FUV bandpass (1010 – 1270 and 1300 – 1565 \AA) of any cool star. We combine flight data with stellar models and archival observations to develop the first SED of a mid-F star. We model the response of a modern Earth-like exoplanet's upper atmosphere to the heightened X-ray and extreme ultraviolet radiation within the habitable zone of Procyon A. These models indicate that this planet would not experience significant atmospheric escape. We simulate observations of the Ly$\alpha$ transit signal of this exoplanet with the \textit{Hubble Space Telescope} (\textit{HST}) and the \textit{Habitable Worlds Observatory} (\textit{HWO}). While marginally detectable with \textit{HST}, we find that \ion{H}{1} Ly$\alpha$ transits of potentially habitable exoplanets orbiting high radial velocity F-type stars could be observed with \textit{HWO} for targets up to 150 pc away.

\vspace{0.8cm}
\end{abstract}

\section{Introduction} \label{s:intro} 

With the successful deployment of the \textit{James Webb Space Telescope} (\textit{JWST}), astronomy has entered the era of detailed characterization of exoplanetary atmospheres. To date there are over 5000 confirmed exoplanets, and nearly 100 \textit{JWST} observing programs focused on exoplanetary studies. It is important to contextualize transmission spectra of exoplanet atmospheres with the energy inputs of their host star \citep{2018haex.bookE..57M,2022arXiv221110490T}. The differing spectral energy distributions (SEDs) of F-, G-, K-, and M-type dwarf stars result in varied photochemical reactions which manifest as diverse atmospheric structures and processes, and correspondingly different signals in transmission spectra \citep{2018haex.bookE..57M}. In order to properly interpret these various spectral features, the radiation environment of the host star must be well characterized.
\par
The types of exoplanets atmospheres that \textit{JWST} is capable of observing are those from exoplanets which orbit close to their host star, and primarily Jupiter and Neptune sized planets. The Transiting Exoplanet Community Early Release Science Program has conducted spectroscopic investigations of three exoplanets: The Hot Saturn WASP-39b, the Hot Jupiter WASP-43b, and another Hot Jupiter WASP-18b \citep{2017jwst.prop.1366B}. These three targets were chosen as \textit{JWST} science and technical demonstrations of cross-referencing instrument performance at overlapping wavelengths (WASP-39b), single visit observations of an entire planetary phase curve (WASP-43b), and secondary eclipse observations with a bright host star (WASP-18b). 
\par
Transmission spectroscopy of WASP-39b revealed the first detections of CO$_2$ and SO$_2$ outside of the Solar System \citep{2022arXiv220811692T,2022arXiv221110490T}. The detection of SO$_2$ absorption at 4.05$\mu$m marked the first direct evidence of photochemistry occurring within an exoplanet atmosphere induced by its G8 host star \citep{2022arXiv221110490T}. While the science results have not yet been released, the data reduction of the first full phase curve study of WASP-43b has been competed and shared with the community to serve as a guide to future phase curve studies in Cycle 2 and beyond \citep{2023arXiv230106350B}. The eclipse observations of WASP-18b have been used to generate thermal emission spectra and 3D mapping of its dayside atmosphere \citep{2023AAS...24112405C,2023AAS...24115905C}. Another notable \textit{JWST} result has been the nondetection of CH$_4$ in the transmission spectrum of the Hot Saturn HAT-P-18b. Photochemistry induced by its K2 V host star may be responsible for photodissociating CH$_4$ high in the exoplanetary atmosphere, however further investigations are needed to confirm this \citep{2022ApJ...940L..35F}.
\par 
Looking towards the future, characterization of FGKM host stars will be critical for modeling and interpreting transmission spectra of the terrestrial atmospheres expected to be observed with the \textit{Habitable Worlds Observatory} (\textit{HWO}) \citep{2019arXiv191206219T,2020arXiv200106683G}. Photochemistry in exoplanet atmospheres is primarily driven by far ultraviolet (FUV; 912 - 1700 \AA) emission from the host star. \ion{H}{1} Lyman alpha (Ly$\alpha$, 1216 \AA) dominates the total FUV emission of late-type stars. Other notable emission lines include \ion{O}{6} (1032 and 1038 \AA), \ion{C}{3} (1174 - 1176 \AA), \ion{Si}{3} (1206 \AA), \ion{N}{5} (1239 and 1243 \AA), \ion{O}{1} (1302, 1305, and 1306 \AA), \ion{C}{2} (1334 - 1336 \AA), \ion{Si}{4} (1394 and 1403 \AA), and \ion{C}{4} (1548 and 1551 \AA). Observations of these emission lines are necessary for a full accounting of FUV driven photochemistry. Stellar emission varies on timescales of minutes to hours during stellar flares, and emission lines are known to behave differently from one another temporally \citep{2018ApJ...867...71L,2020AJ....160..237F}. The flare frequency of F-type stars is not well characterized, making simultaneous observations of several FUV emission lines important for properly modeling and interpreting transmission spectra.
\par
The stellar UV spectrum (which varies with spectral type) combined with exoplanetary atmospheric composition dictates the types of photochemical reactions that take place within an exoplanet's atmosphere. One example is the interpretation of O$_2$ features as a biosignature in transmission spectra. The ratio of FUV to near ultraviolet (NUV 1700 - 4000 \AA) light regulates the production of abiotic O$_2$ and O$_3$ within terrestrial atmospheres. FUV emission dissociates atmospheric CO$_2$ and produces O$_2$, whereas NUV emission dissociates O$_2$. \cite{2015ApJ...812..137H} finds that the FUV to NUV ratio of F and G dwarfs is such that an appreciable concentration of abiotic O$_2$ cannot be maintained. In this instance, a strong detection of O$_2$ may be a true biosignature. The decreased NUV continuum emission of cooler K and M dwarfs results in less O$_2$ dissociation, making an abiotic buildup of O$_2$ possible and potentially misinterpreting this as a biosignature. \cite{2018ApJ...854...19R} and \cite{2014E&PSL.385...22T} describe a similar effect observed for O$_3$ in terrestrial atmospheres. The transmission spectrum must be put into context through the photochemistry induced by the host star.
\par
In addition to photochemical effects, the X-ray and EUV (XUV; 10 - 1180 \AA) emission of a host star is the critical driver of atmospheric escape. Within planetary atmospheres lies the thermosphere, a layer which is heated by the absorption of incoming XUV radiation from the host star \citep[See][]{2005JGRA..110.1312W}. Above the thermosphere lies the exosphere, the outermost layer of planetary atmospheres, whose nearly constant temperature is determined by thermospheric heating. The interface between the exosphere and the thermosphere, known as the exobase, is the altitude at which particles become collisionless. It is at the exobase where light particles such as H$^0$ can escape to space if their velocity exceeds the escape velocity of the planet. This atmospheric escape mechanism is known as Jeans escape.
\par
Jeans escape is characterized by the Jeans parameter $\lambda_J$, the ratio of gravitational energy to thermal energy of an atmospheric species. With sufficient XUV heating, thermal energy can overcome gravitational energy and hydrodynamic escape, or blowoff, is possible. The blowoff state can result in significant atmospheric loss, and has been observed on a handful of exoplanets \citep[e.g.][]{2004ApJ...604L..69V,2018A&A...620A.147B,2023AJ....165...62Z}. The blowoff state begins when $\lambda_J$ $<$ 1.5 \citep{2008JGRE..113.5008T}. With knowledge of the incoming XUV emission from the host star in conjunction with a planetary atmosphere model, the atmospheric escape of exoplanets can be estimated.
\par
Exoplanet host star SEDs, which cover the full stellar emission range from X-rays to the infrared (IR), are utilized to model and interpret the interactions between host stars and the atmospheres of their exoplanets \citep[e.g.][]{2015ApJ...812..137H,2022MNRAS.512.4877B}. The SED of one star can be used as proxies for other stars of a similar spectral type \citep[e.g.][]{2022MNRAS.512.4877B,2022ApJ...936..177Y}. The Measurements of the Ultraviolet Spectral Characteristics of Low-mass Exoplanetary Systems (MUSCLES) and subsequent Mega-MUSCLES Treasury Surveys developed SEDs for a multitude of K and M dwarf stars \citep{2016ApJ...820...89F,2016ApJ...824..101Y,2016ApJ...824..102L,2019ApJ...871L..26F,2021ApJ...911...18W}. The SEDs developed through the MUSCLES family have been used in a variety of ways. Their primary use is as accurate stellar inputs of K and M dwarfs for exoplanet atmosphere models \citep[e.g.,][]{2022MNRAS.512.4877B,2022MNRAS.513.6125N,2021ApJ...921...27H}, however, the SEDs also see use including the development of extreme ultraviolet (EUV; 100 - 912 \AA) stellar reconstruction techniques \citep{2021ApJ...913...40D}, and spectral evolution of K and M dwarfs \citep{2014AJ....148...64S,2021ApJ...907...91L,2022ApJ...929..169R}.
\par
Our Sun is the primary proxy for the radiation environment around G-type stars \citep[e.g.][]{2009GeoRL..36.1101W,2012ApJ...757...95C}. It is the only star which has been observed in all wavelength regimes, particularly in the EUV. For all other stars, the lack of an astronomical EUV facility as well as \ion{H}{1} attenuation from the interstellar medium (ISM) from $\sim$400 - 912 \AA\ makes empirical SED creation difficult for planet hosting stars \citep{2022JATIS...8a4006F}. Because the SED of our Sun does not suffer from this attenuation, it is an excellent proxy for similar G dwarfs, such as WASP-39 \citep{2022arXiv221110490T}.
\par
Unlike G, K, and M dwarfs, SEDs for F-type stars are not currently available in the literature. Studies in the past have either not utilized the full spectral range \citep[e.g.][]{2005ApJ...633..424A,2012A&A...540A...5C,2015ApJ...812..137H}, or combined observations of F-type stars with scaled solar spectra \citep[e.g.][]{2022MNRAS.512.4877B}. Spectral coverage below $\sim$1200 \AA\ is generally missing in the SEDs of mid-F stars. In addition, the observability of terrestrial \ion{H}{1} exospheres has been explored for planets around nearby M dwarfs and solar analogs \citep{2018ExA....45..147C,2019A&A...622A..46D,2019A&A...623A.131K}, but observability for F-type host stars is currently unknown.
\par
We address the need for a broadband FUV observation of a main sequence mid-F star with the Suborbital Imaging Spectrograph for Transition-region Irradiance from Nearby Exoplanet host stars (SISTINE) sounding rocket payload. In Section \ref{s:sistine} we describe the SISTINE instrument and the results from its observation of Procyon A. Section \ref{s:SED} details the development of the full SED of an F-type star using the flight data. The SED is used in Section \ref{s:exoplanets} to predict the impact of an F-type star on the \ion{H}{1} exosphere of an Earth-like exoplanet.

\section{The SISTINE Sounding Rocket Payload} \label{s:sistine} 

The SISTINE sounding rocket payload was developed to characterize the radiation environments of exoplanet host stars \citep{2016SPIE.9905E..0AF,2016JAI.....540001F}. The instrument was designed to capture several FUV emission features of FGKM dwarf stars over a broad spectral bandpass (1010 - 1270 and 1300 - 1565 \AA), as emission features from these types of stars are known to independently vary on the timescales of a sounding rocket flight, even for inactive stars \citep{2018ApJ...867...71L,2020AJ....160..237F}. The instrument design also acts as a pathfinder for a future broadband FUV spectrograph on board \textit{HWO} \citep{2016JATIS...2d1203F,2017SPIE10397E..13F}. 
\par
SISTINE has a spectral resolving power of R$\sim$1500, capable of resolving several key emission features from \ion{O}{6} (1032 \AA) to \ion{C}{4} (1551 \AA). The imaging capabilities of SISTINE allow for the resolution of objects with a separation greater than $\sim$4'', such as binary star systems. SISTINE achieves these resolutions through its instrument design and with the implementation of several novel FUV technologies. These enabling technologies are tested on SISTINE in preparation for future large IR/O/UV missions as called for by the Astro2020 Decadal Survey \citep{Astro2020}, such as \textit{HWO}.
\par
For the second flight of SISTINE (SISTINE-2), we observed the Procyon A+B binary system. Procyon A (F5 IV-V \citep{2013yCat....1.2023S}) is the main science target of SISTINE-2, with the objective of capturing its FUV radiation. Its companion, Procyon B (DQZ, \citealt{2013yCat....1.2023S}), was a secondary non-critical science target. At the time of launch, the two stars had a separation of ~4.9'' making the binary pair resolvable by SISTINE-2. While the Procyon A+B system is not currently known to host exoplanets, Procyon A acts as, by far, the most observationally favorable representative of other mid-F host stars. A brief description of the instrument design and predicted performance is given in Section \ref{ss:buildup}. Further details are found in \cite{2021SPIE11821E..0HC}. The SISTINE-2 flight results are presented in Section \ref{ss:flight}.

\begin{figure}[!ht]
\epsscale{1.15}
\plotone{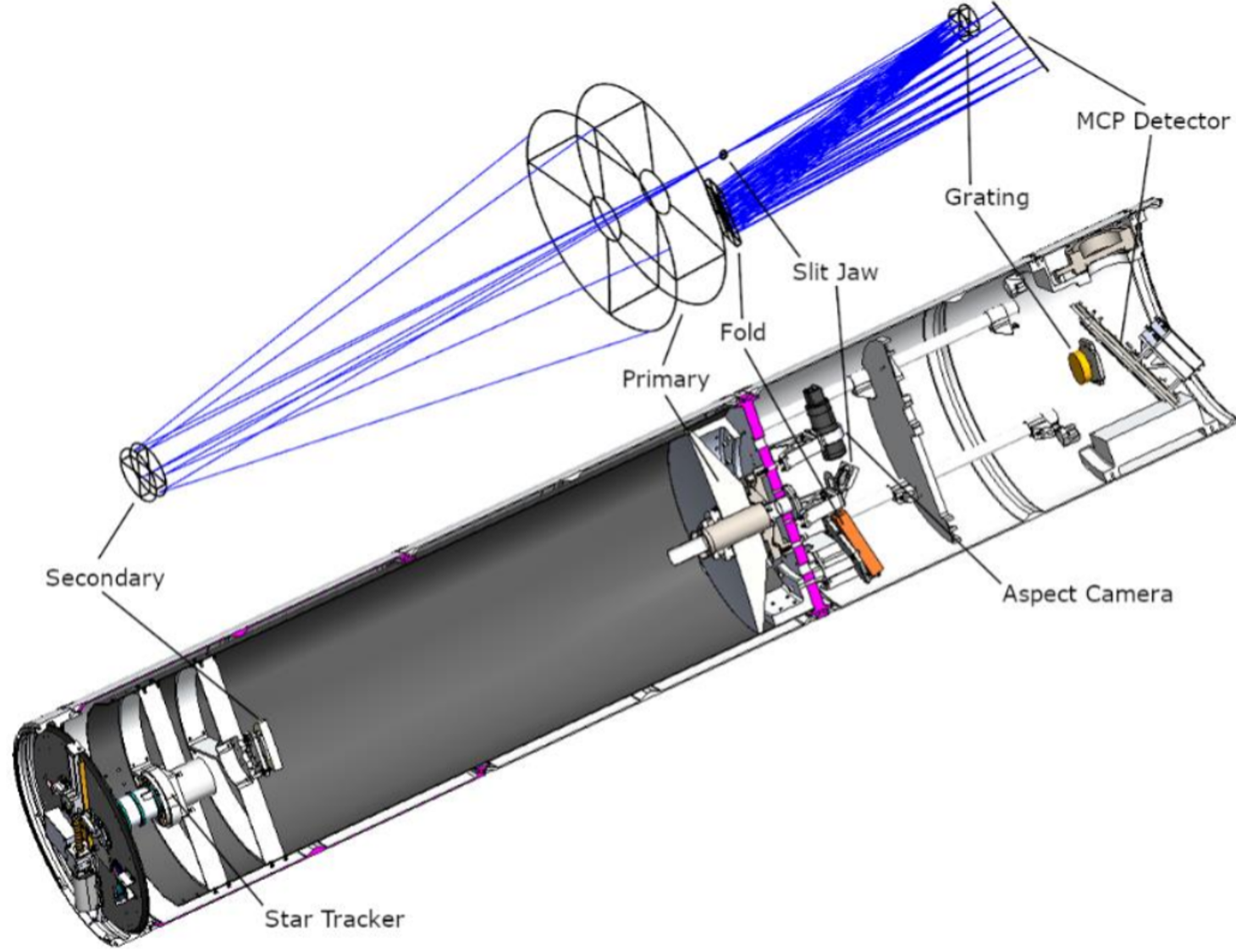}
\caption{The optical ray trace and mechanical layout of the SISTINE-2 payload. Light enters the instrument from the lower left, reflects off the primary mirror to the secondary mirror, and is focused at the mirrored slit jaw. The aspect camera images the slit jaw in real time during flight. Light that passes through the slit is diffracted off the grating towards the fold mirror, and the spectrum is imaged onto the MCP detector.}
\label{fig:optomech} 
\end{figure}

\subsection{Assembly and Calibration} \label{ss:buildup} 

The SISTINE instrument comprises four reflective optics and a microchannel plate (MCP) detector, divided into an \textit{f}/14 Cassegrain telescope and a 2.1$\times$ magnifying FUV imaging spectrograph, resulting in a total instrument focal ratio of \textit{f}/30 \citep{2016SPIE.9905E..0AF,2016JAI.....540001F,2021SPIE11821E..0HC}. The optomechanical design of SISTINE is shown in Figure \ref{fig:optomech}. 
\par
The telescope section consists of a primary and secondary mirror, which images light onto a slit jaw. Light that does not pass through the slit is reflected and reimaged by an aspect camera to monitor the pointing of the instrument during flight. The primary and secondary mirrors employ enhanced lithium fluoride (eLiF) coatings. These coatings improve FUV reflectivity compared to traditional LiF by heating the substrate as it is coated, enabling a higher packing density in the LiF layer \citep{2014SPIE.9144E..4GQ,2017ApOpt..56.9941F}. With a 0.5 m diameter, the SISTINE primary mirror is one of the largest optics to be utilized in a sounding rocket payload, and is the largest to be coated with eLiF. SISTINE is the first instrument to test eLiF coatings in space. The secondary mirror includes an additional aluminum trifluoride (AlF$_3$) capping layer, applied using atomic layer deposition (ALD) to protect the optic by applying a buffer layer for humidity induced degradation of the hygroscopic LiF layer \citep{2016JATIS...2d1206H,2017ApOpt..56.9941F}. ALD capping layers are tested in space for the first time with SISTINE.
\par
The spectrograph consists of a blazed, holographically ruled reflective grating, a rectangular fold mirror which is powered along its long axis, and two identical MCP detector segments. The grating is coated with LiF due to the fact that it is a replica, and the fold mirror is coated with eLiF. The MCP detector plates are ALD processed, which enables the use of stronger substrate materials, improved temporal stability, and larger detector areas \citep{2012NIMPA.695..168S,2015SPIE.9601E..0SE,2016JVSTA..34aA128O}. 

\subsection{SISTINE-2 Flight Results} \label{ss:flight} 

SISTINE-2 launched on November 8\textsuperscript{th} 2021 at 9:25 UTC from White Sands Missile Range (WSMR) to observe the Procyon A+B binary star system. SISTINE-2 remained on target for 339 s. During the observation, instrumental defocus was apparent in the data and aspect camera. Despite this defocus, the FUV spectrum of Procyon A was successfully observed at a high enough resolution and sensitivity to capture all critical stellar emission features targeted by the mission. The flux calibration and characterization of the observed defocus is described below.

The FUV spectrum of Procyon B was not detected, however this was not a consequence of the defocus observed during flight. We examine the FUV spectra of two nearby white dwarfs of similar spectral type and temperature as Procyon B (DQZ, 7740 $\pm$ 50 K \citep{2002ApJ...568..324P}); van Maanen 2 (DZ, 6130 $\pm$ 110 K \citep{2017AJ....154...32S}) and GJ 440 (DQ, 8490 $\pm$ 270 K \citep{2001ApJS..133..413B}). These two white dwarfs were observed with the Cosmic Origins Spectrograph (COS) on board the \textit{Hubble Space Telescope} (\textit{HST}), using the G140L grating. In a similar manner as presented in \cite{2021SPIE11821E..0HC}, we use these archival spectra to make a prediction on the observability of Procyon B. We find that even in focus, SISTINE-2 likely would not have been able to observe the FUV radiation from Procyon B due to a low signal to noise ratio (SNR) achievable within the duration of a typical sounding rocket flight.

\subsubsection{Characterizing the Defocus} \label{sss:defocus} 

\begin{figure}[!ht]
\epsscale{1.15}
\plotone{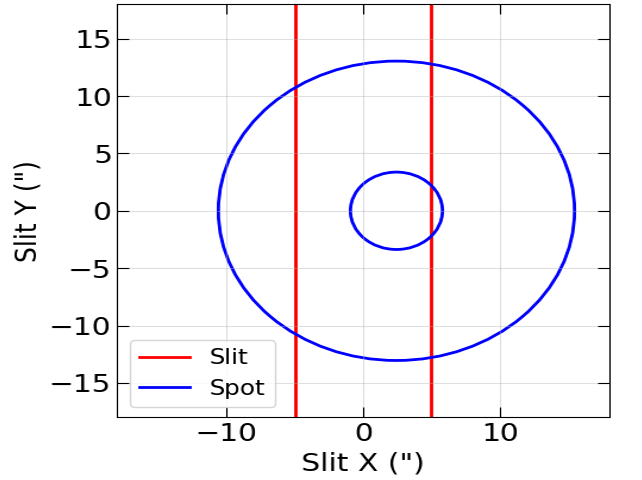}
\caption{A diagram of the out of focus SISTINE-2 spot projected onto the instrument slit. The instrument slit is bounded by the parallel red lines. The SISTINE-2 spot is bounded by the concentric blue circles. Approximately 43\% of starlight is transmitted into the spectrograph.}
\label{fig:donut} 
\end{figure}

We first characterize the defocus of the optical system observed during flight. After recovery of the payload, focus checks of the telescope section in the field revealed that the SISTINE-2 focus was comparable to the measured focus prior to launch. This suggested that a thermal effect during flight was the cause of the telescope defocus. We confirmed this in a laboratory test in which the primary mirror bulkhead is heated to the temperature observed during flight ($\sim$45$^{\circ}$ C), which resulted in a defocused telescope. 
\par
We estimate the amount of light that did not pass through the slit by using two concentric circles to simulate the defocused spot, shown in Figure \ref{fig:donut}. By fitting to observed emission features on the 2D detector data, the fraction of light that did not pass through the slit is estimated. We find that $\sim$43\% of light collected by the telescope was transmitted into the spectrograph. The thermal defocus was successfully mitigated in preparation for the third flight of SISTINE (SISTINE-3), and will be described in Nell et al. (in prep. 2023). The science results of SISTINE-3 (launched July 6\textsuperscript{th} 2022 at 13:47 UTC) will be described in Behr et al. (in prep. 2023).   

\subsubsection{Background Subtraction} \label{sss:background} 

\begin{figure*}[!ht]
\epsscale{1.18}
\plotone{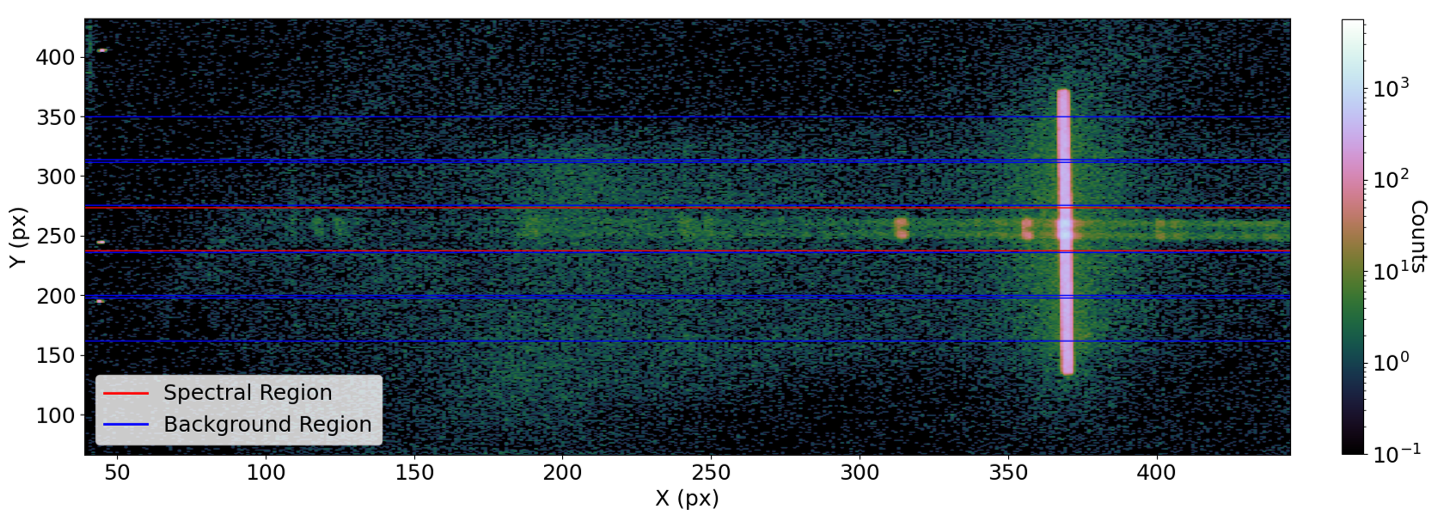}
\caption{The SISTINE-2 short wavelength flight image of Procyon A. The spectral extraction region is shown in red, with four background regions, shown in blue, above and below the spectral region. Geocoronal Ly$\alpha$ fully illuminates the instrument slit at pixel $\sim$370 along the X axis. The image is binned by 16 to best juxtapose the stellar spectrum, geocoronal Ly$\alpha$ emission, and the background.}
\label{fig:bgSub} 
\end{figure*}

With the characterization of the defocus experienced during flight, a flux calibration is performed on the spectrum. The detector's X axis records spectral data, while its Y axis (aligned with the instrument slit image) records spatial data. We begin with a background subtraction, where we treat the two detector segments independently and perform the following procedure on each segment. A rectangular extraction region is defined which captures the spectral trace of Procyon A. Four background regions of equal size to the spectral region are defined, two above and two below the spectral region. Figure \ref{fig:bgSub} presents these regions on the short wavelength detector segment image (1010 - 1270 \AA) taken during flight.
\par
Due to relative rotations between the instrument slit, grating, and detector, the instrument image is not aligned with the detector. To account for this in our background subtraction, we perform a cross correlation on the Ly$\alpha$ geocoronal (airglow) emission to align the spectral and background regions. The relatively bright Ly$\alpha$ airglow feature fully illuminates the slit, enabling the cross correlation for the four background regions. Ly$\alpha$ airglow is only present on the short wavelength detector segment, and we apply the same shifts to the background spectra on the long wavelength detector segment. The spectral and background regions are collapsed into 1D spectra, the four background spectra are averaged into a single background spectrum. We subtract the average background spectrum from the collapsed spectral region to produce the background subtracted 1D spectrum of Procyon A.

\subsubsection{Count Rates} \label{sss:counts} 

\begin{figure}[!ht]
\epsscale{1.18}
\plotone{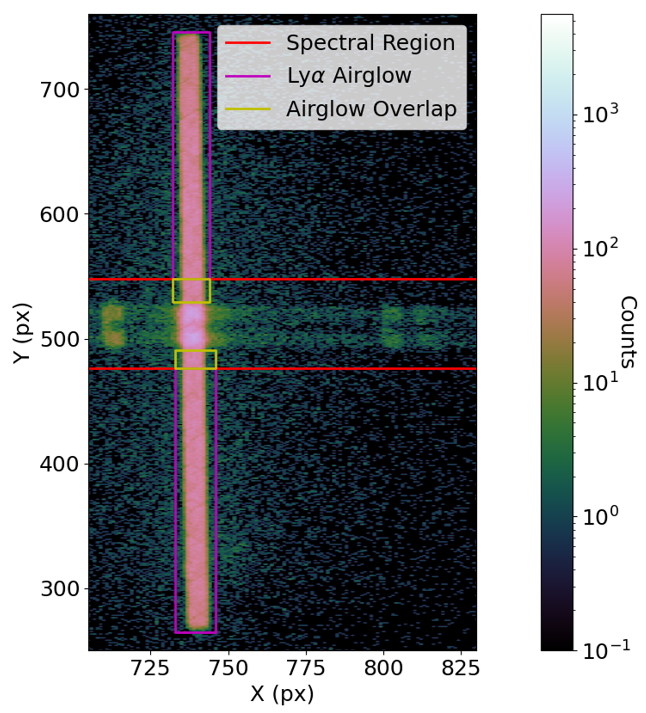}
\caption{The flight image in the region about Ly$\alpha$ airglow. As with Figure \ref{fig:bgSub}, the red box bounds the spectral region. The airglow subregions within the spectral region are shown in yellow, and the remaining airglow is bounded by a purple box. To the left of Ly$\alpha$ is \ion{Si}{3} emission and to the right is emission from the \ion{N}{5} doublet. This image is binned by 8 to best contrast Ly$\alpha$ airglow from the stellar spectrum.}
\label{fig:agBox} 
\end{figure}

From the 2D detector data, we calculate count rates for the spectrum, and the background, with a separate treatment of Ly$\alpha$ airglow. Within the spectral extraction region, we define two rectangular subregions which encapsulate Ly$\alpha$ airglow, and counts within this region are subtracted from the spectral region. The average Procyon A count rate over the entire instrument bandpass is 344 counts s$^{-1}$. We define two rectangular airglow regions above and below the spectral trace which encompass the Ly$\alpha$ airglow. The counts in these regions are added to the two airglow subregions within the spectral region and a final average airglow count rate is calculated to be 622 counts s$^{-1}$. See Figure \ref{fig:agBox} for a visualization of these Ly$\alpha$ airglow regions. The background count rate is then determined by subtracting the spectral and airglow count rates from the global count rate. We find an average background count rate of 762 counts s$^{-1}$ integrated over the $\sim$9 cm$^2$ detector face ($\sim$85 counts cm$^{-2}$ s$^{-1}$). We find the stellar and airglow count rate to be stable throughout the 339s exposure. The background count rate decreases in the first minute of the exposure before settling to a stable count rate throughout the remainder of the exposure.

\subsubsection{Flux Calibration} \label{sss:fluxcal} 

Using the background subtracted spectrum developed in Section \ref{sss:background}, we perform a flux calibration of the spectrum. A wavelength solution is calculated for the in-flight spectrum using known wavelengths of bright emission features. We find that this wavelength solution matches those developed from laboratory calibrations. We use this wavelength solution, the in-flight exposure time, and the component level effective area presented in \citep{2021SPIE11821E..0HC} to produce the flux calibrated spectrum of Procyon A. 
\par
Due to the observed defocus during flight, we recalculate the in-flight resolution of the instrument. We use the isolated \ion{Si}{3} emission line, and fit a boxcar to the emission profile, which approximated the flat topped emission features in the 1D spectrum. We find that the in-flight spectral resolution is reduced to R$\sim$500, consistent with the SISTINE-2 slit limiting the instrument resolving power. This resolving power was sufficient for isolating all spectral features of interest in the stellar data. 
\par
The UV variability of F-type stars on the timescale of decades is not well characterized. We make a variability assessment by utilizing the available observations of Procyon A taken with the \textit{International Ultraviolet Explorer} (\textit{IUE}). To quantify the temporal variability of the star on several year timescales, we employ a process similar to what is described in \cite{1991ApJ...375..704A}. We utilize all available short wavelength (1150 - 2000 \AA) observations made with the Short Wavelength Prime (SWP) camera, at R$\sim$200. Integrated fluxes are calculated for the entire bandpass (excluding the airglow contaminated Ly$\alpha$ line), and for few relatively bright emission lines such as the \ion{C}{2} multiplet (1334 - 1336 \AA). We find that the stellar variability is low ($\sim$3\% for broadband emission and $\sim$4-9\% for individual emission features) for \textit{IUE} observations made between 1979 and 1990 with the SWP camera. We note that this analysis does not cover the long term variability of all individual emission features of interest, which are observed to vary independently from one another in solar observations \citep{Tom_Woods2020-pj}.
\par
Observations of Procyon A taken with the Space Telescope Imaging Spectrograph (STIS) on board \textit{HST} and the \textit{Far Ultraviolet Spectroscopic Explorer} (\textit{FUSE}) were used to make a comparison between the archival observations and the sounding rocket observation. Based on the broadband flux and bright emission line fluxes in the \textit{IUE} bandpass, the \textit{FUSE} (2001) and the STIS (2011) are not likely to display large deviations in flux from the SISTINE-2 observation and can be compared. We convolve the \textit{FUSE} and STIS data with the boxcar SISTINE-2 Line Spread Function (LSF) estimated from the \ion{Si}{3} emission line. When comparing the SISTINE-2 data to the archival data, the fluxes of individual emission features and the stellar continuum were found to be lower than what is observed in the STIS and \textit{FUSE} data. These discrepancies are too large to be attributed to stellar variability, and it is more likely to be caused by a decrease in instrument throughput.

\begin{figure*}[!ht]
\epsscale{1.117}
\plotone{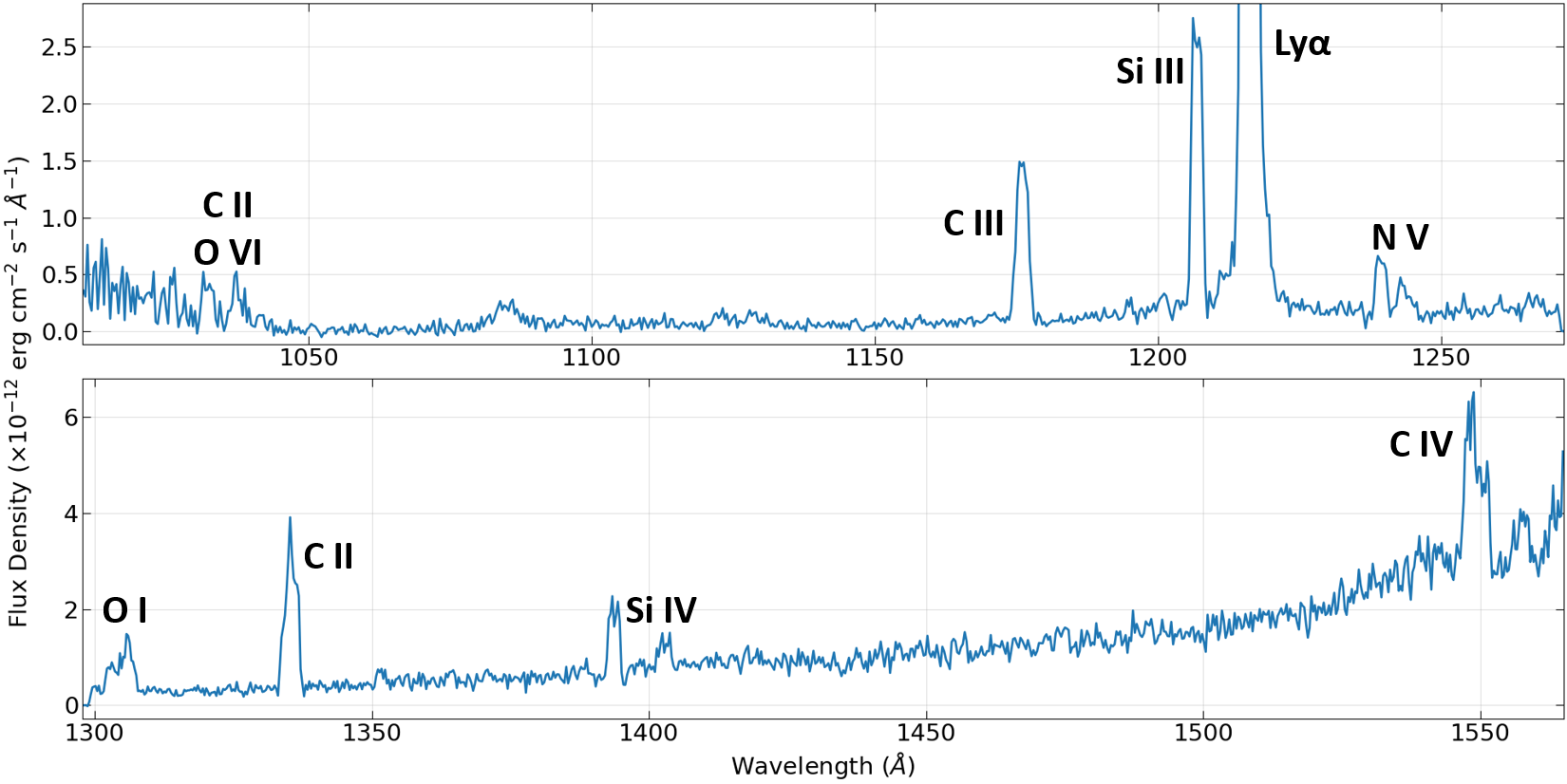}
\caption{The FUV spectrum of Procyon A observed by SISTINE-2. The flight spectrum is binned by 8 to highlight key spectral features while maintaining spectral resolution. The short wavelength spectrum is plotted above the long wavelength spectrum, with key features labeled. ISM attenuated Ly$\alpha$ emission peaks at $\sim$2.6$\times$10$^{-11}$ erg cm$^{-2}$ s$^{-1}$ \AA$^{-1}$.}
\label{fig:flightSpec} 
\end{figure*}

To reconcile differences between the sounding rocket and archival spectra, we calculate integrated fluxes for bright emission features in the SISTINE-2 and archival spectral bandpasses. For each individual detector segment, we calculate a single scale factor for the SISTINE-2 effective area which best reconciles the observed differences. We find that the effective area on the short wavelength detector segment was $\sim$70\% of the expected value (after accounting for the drop due to defocus), and the long wavelength segment was $\sim$54\% of the expected value. These decreases in effective area are reasonably attributable to overall degradation of instrument throughput associated with extended environmental exposure to individual components during integration field operations \citep{2018arXiv180103102F}.

\subsubsection{Emission Line Fluxes} \label{sss:lineflux} 

\begin{deluxetable*}{cccccc}[!ht]
\tablenum{1} 
\tablecaption{Integrated flux measurements of FUV emission lines of Procyon A. Reported Ly$\alpha$ emission is airglow subtracted, but remains attenuated by the ISM. Solar fluxes during solar minimum and maximum are shown for comparison as would be observed if the Sun were at the same distance as Procyon A \citep{2020SpWea..1802588C}. Archival line fluxes calculated from \textit{FUSE} and STIS data are also shown for comparison, where available. Solar Ly$\alpha$ fluxes are omitted, as they cannot be compared to the ISM attenuated SISTINE-2 and STIS profiles. \label{tab:emission}}
\tablewidth{0pt} 
\tablehead{
\colhead{Emission Line} & \colhead{Integrated Flux} &
\colhead{Solar Min. Flux} &
\colhead{Solar Max. Flux} & \colhead{\textit{FUSE} Flux} & \colhead{STIS Flux} \\
\colhead{(\AA)} & \colhead{(erg cm$^{-2}$ s$^{-1}$)} & \colhead{(erg cm$^{-2}$ s$^{-1}$)} & \colhead{(erg cm$^{-2}$ s$^{-1}$)} & \colhead{(erg cm$^{-2}$ s$^{-1}$)} & \colhead{(erg cm$^{-2}$ s$^{-1}$)} 
}
\startdata
\ion{O}{6}, \ion{C}{2} & \multirow{2}*{$1.88^{+0.55}_{-0.57} \times 10^{-12}$} & \multirow{2}*{$1.75 \times 10^{-13}$} & \multirow{2}*{$2.46 \times 10^{-13}$} & \multirow{2}*{$2.54 \pm 0.04 \times 10^{-12}$} & \multirow{2}*{---} \\ 
(1032 - 1038) & & & & & \\ 
\ion{C}{3} & \multirow{2}*{$3.61^{+0.31}_{-0.35} \times 10^{-12}$} & \multirow{2}*{$1.55 \times 10^{-13}$} & \multirow{2}*{$1.89 \times 10^{-13}$} & \multirow{2}*{$1.51 \pm 0.03 \times 10^{-12}$} & \multirow{2}*{$4.25 \pm 0.07 \times 10^{-12}$} \\ 
(1174 - 1176) & & & & & \\ 
\ion{Si}{3} & \multirow{2}*{$5.31^{+0.52}_{-0.48} \times 10^{-12}$} & \multirow{2}*{$2.04 \times 10^{-13}$} & \multirow{2}*{$3.01 \times 10^{-13}$} & \multirow{2}*{---} & \multirow{2}*{$4.84 \pm 0.03 \times 10^{-12}$} \\ 
(1206) & & & & & \\ 
Ly$\alpha$ & \multirow{2}*{$6.73^{+0.35}_{-0.34} \times 10^{-11}$} & \multirow{2}*{---} & \multirow{2}*{---} & \multirow{2}*{---} & \multirow{2}*{$5.53 \pm 0.01 \times 10^{-11}$} \\ 
(1216) & & & & & \\ 
\ion{N}{5} & \multirow{2}*{$1.93^{+0.35}_{-0.36} \times 10^{-12}$} & \multirow{2}*{$9.82 \times 10^{-14}$} & \multirow{2}*{$1.21 \times 10^{-13}$} & \multirow{2}*{---} & \multirow{2}*{$1.67 \pm 0.02 \times 10^{-12}$} \\ 
(1239, 1243) & & & & & \\
\ion{C}{2} & \multirow{2}*{$7.60^{+0.69}_{-0.70} \times 10^{-12}$} & \multirow{2}*{$3.03 \times 10^{-13}$} & \multirow{2}*{$4.08 \times 10^{-13}$} & \multirow{2}*{---} & \multirow{2}*{$8.48 \pm 0.02 \times 10^{-12}$} \\ 
(1334 - 1336) & & & & & \\
\ion{Si}{4} & \multirow{2}*{$4.90^{+1.02}_{-0.97} \times 10^{-12}$} & \multirow{2}*{$2.24 \times 10^{-13}$} & \multirow{2}*{$2.96 \times 10^{-13}$} & \multirow{2}*{---} & \multirow{2}*{$4.43 \pm 0.05 \times 10^{-12}$} \\ 
(1394, 1403) & & & & & \\
\ion{C}{4} & \multirow{2}*{$1.15^{+0.18}_{-0.15} \times 10^{-11}$} & \multirow{2}*{$4.17 \times 10^{-13}$} & \multirow{2}*{$5.07 \times 10^{-13}$} & \multirow{2}*{---} & \multirow{2}*{$9.64 \pm 0.11 \times 10^{-12}$} \\ 
(1548, 1551) & & & & & \\
\enddata
\end{deluxetable*}

The flux calibrated spectrum of Procyon A is presented in Figure \ref{fig:flightSpec}, accounting for the instrument defocus and effective area degradation. We list a collection of emission line fluxes measured with flux calibrated SISTINE-2 data in Table \ref{tab:emission}, with flux from the stellar continuum subtracted. We do not include \ion{O}{1} emission in Table \ref{tab:emission} due to the emission features being located on the edge of the long wavelength detector segment. \ion{O}{6} emission is contaminated with unresolved emission from \ion{C}{2} (1036 and 1037 \AA), and so we report the combined flux originating from these two species in Table \ref{tab:emission}. 
\par
The Ly$\alpha$ airglow emission observed during flight is also calculated. In Section \ref{sss:counts}, we determined that the average Ly$\alpha$ airglow count rate was 622 counts s$^{-1}$ during flight. Assuming the $\sim$54\% drop in spectrograph throughput due to the defocus, and a further drop of $\sim$30\% from instrument throughput degradation, the observed geocoronal Ly$\alpha$ emission is consistent with $\sim$2000 Rayleighs, the expected value during a night observation \citep{2022cosi.book...14J,2022stii.book...21P}.
\par
For comparison, Table \ref{tab:emission} additionally includes solar fluxes obtained from the Flare Irradiance Spectral Model - Version 2 (FISM2), scaled to the distance to Procyon A. We include fluxes modeled during solar minimum and solar maximum. The FUV emission lines of Procyon A as observed by SISTINE-2 are notably stronger than the Sun's, during both solar minimum and maximum. As will be explored in Section \ref{s:exoplanets}, the habitable zones of stars similar to Procyon A are generally further out than the Sun's habitable zone; as a result, orbiting exoplanets within this region do not experience significantly stronger FUV fluxes from their host star.
\par
Table \ref{tab:emission} also includes measured line fluxes from the archival \textit{FUSE} and STIS FUV observations of Procyon A. Of the emission lines of interest, \textit{FUSE} data contains the \ion{O}{6} + \ion{C}{2} and \ion{C}{3} features, while STIS data contains all but the \ion{O}{6} + \ion{C}{2} features. There is rough general agreement between the novel and archival line fluxes. Of these emission features, only \ion{C}{3} was captured by all three instruments. The SISTINE-2 \ion{C}{3} flux is bounded between the measured \textit{FUSE} and STIS fluxes for the same emission feature. These differences in emission line strength from data taken in 2001 (\textit{FUSE}), 2011 (STIS) and 2021 (SISTINE-2) may be related to changes throughout the stellar activity cycle of Procyon A; a better understanding of stellar variability on these timescales may confirm this to be the case.

\section{The Full SED of Procyon A} \label{s:SED} 

\begin{figure*}[!ht]
\epsscale{1.15}
\plotone{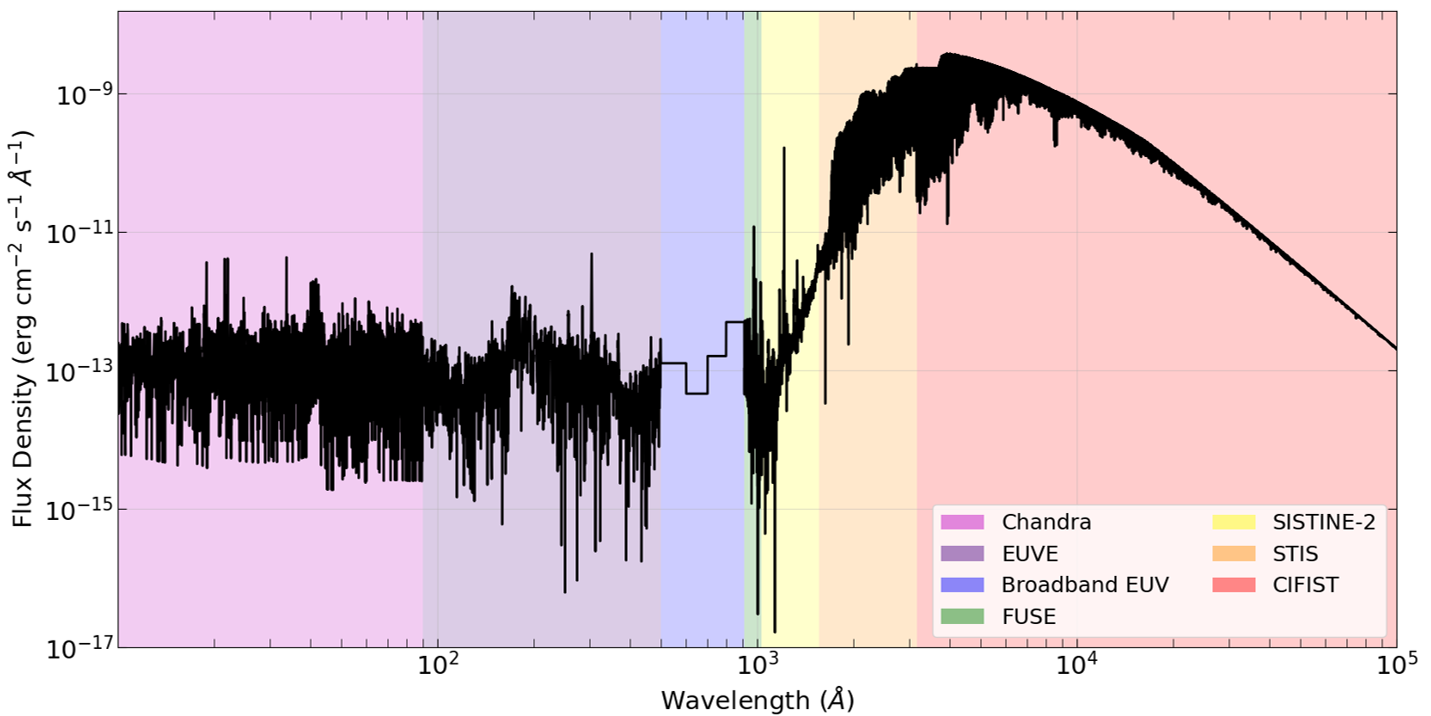}
\caption{The full SED of Procyon A as observed from Earth. Colored regions represent the different observations or models which were combined to produce the SED, each described in detail throughout Section \ref{s:SED}.}
\label{fig:SED} 
\end{figure*}

The observed FUV spectrum of Procyon A developed in Section \ref{s:sistine} is combined with archival observations and models of Procyon A to develop the first X-ray through IR SED (10 \AA\ -- 1 mm) of a mid F-type star. We present the full SED in Figure \ref{fig:SED}. The individual components are described in the following subsections, and are denoted in Figure \ref{fig:SED} by shaded regions of different colors. Some of the data used for the creation of the SED were obtained from the Mikulski Archive for Space Telescopes (MAST) database, which can be accessed via \dataset[DOI: 10.17909/gg3s-cc68]{http://dx.doi.org/10.17909/gg3s-cc68}.

\subsection{Optical and IR: BT-Settl CIFIST Model} \label{ss:oir} 

We use the BT-Settl CIFIST model for the optical and infrared (IR) regimes of Procyon A \citep{2015A&A...577A..42B}. This model component ranges from 3158 \AA\ to 1 mm. The BT-Settl CIFIST models require three input parameters: the effective temperature, surface gravity, and metallicity of the star. We adopt the model grid values of T$_{eff}$ = 6500 K, log(g) = 4.0, and [Fe/H] = 0.0, consistent with the values reported for Procyon A in \cite{2019A&A...624A..19B} of T$_{eff}$ = 6474 $\pm$ 94 K, log(g) = 3.96 $\pm$ 0.07, and [Fe/H] = 0.01 $\pm$ 0.07. The model spectrum is given as a surface flux, and is uniformly scaled to match near ultraviolet (NUV) data taken with STIS in an overlap window of 2658 - 3158 \AA.

\subsection{NUV: STIS ASTRAL} \label{ss:stis} 

We use the STIS ASTRAL (Advanced Spectral Library Project)\footnote{\url{https://archive.stsci.edu/prepds/astral/}} observations of Procyon A to cover the NUV range between 1565 and 3158 \AA, covering the wavelength range between the SISTINE-2 observation and the CFIST model. The ASTRAL program aimed to collect high quality UV spectra of nearby bright stars with the STIS instrument by splicing together echellograms taken with multiple echelle grating modes. The E140H portion of the ASTRAL coadded spectrum was additionally used in the SED to cover the gap in wavelength coverage of SISTINE-2 (1270.5 - 1299.6 \AA) due to the physical gap between the detector segments. This E230H portion is convolved to the spectral resolution of SISTINE-2, and stitched in between the two SISTINE-2 bandpasses.

\subsubsection{Stellar \texorpdfstring{Ly$\alpha$}{Lyalpha} Reconstruction} \label{sss:lya} 

The Ly$\alpha$ reconstruction is performed using the techniques developed in \cite{airglow} (hereafter CA23). Stellar Ly$\alpha$ reconstructions have been demonstrated for spectral resolution as low as R$\sim$1000 \citep{2021ApJ...911..112Y}. Due to the defocus observed during the flight of SISTINE-2, the spectral resolution at Ly$\alpha$ (R$\sim$500) was not high enough to produce a stable and reliable reconstruction of the underlying stellar emission profile within the expected range of reconstruction parameters. For Procyon A, a separate spectrum only using E140M data is additionally available in the ASTRAL library, which we used to conduct a stellar reconstruction. \cite{2018ApJ...867...71L} describes stellar Ly$\alpha$ emission as a ``gentle giant'' as it does not vary as much as other FUV emission features within the same timescale, motivating the use of the STIS Ly$\alpha$ profile for the reconstruction. In the wavelength range of 1213.9 - 1217.4 \AA, we recover the stellar Ly$\alpha$ profile of Procyon A and replace the SISTINE-2 data in this region within the SED.
\par
We simultaneously model the underlying stellar emission and attenuation from the interstellar medium (ISM), and the model is convolved to the STIS E140M resolution. We assume the stellar Ly$\alpha$ profile displays self-reversal centered on the line core, as observed on the Sun \citep[e.g.][]{2012A&A...542L..25L}. Further details on model components and techniques can be found in Section 3 of CA23. When fitting the model to the E140M data, we exclude the central region of the emission line; the ISM completely attenuates the line core, leaving behind residual airglow from the STIS background subtraction routine. Removing this residual airglow allows the model to come to a reasonable fit to the attenuated stellar emission. 
\par
For reconstructing the stellar emission of Procyon A, we constrain the model's search range of the \ion{H}{1} ISM column density. In CA23, the model searches through a $log_{10}(N_{col,HI})$ range of 17.1 - 19.1, to accommodate for F-, G-. K-, and M-type dwarfs at various distances. We instead tailor the model towards this individual star, based on the reported value of 18.1 in \cite{2005ApJS..159..118W} for Procyon A.
\par
We add an additional attenuating component to our Ly$\alpha$ model of Procyon A. Due to our proximity to Procyon A, the attenuating effects of its astrosphere (the heliosphere equivalents of other stars) can be observed in the stellar profile. For Procyon A, modeling the stellar emission and ISM attenuation alone cannot reproduce the observed E140M data, whereas the inclusion of astrospheric absorption results in good agreement \citep{2005ApJS..159..118W}. We set the initial values of the astrospheric column density, radial velocity, and Doppler b parameter to 10$^{14.65}$ cm$^{-2}$, 0.0 km s$^{-1}$, and 25 km s$^{-1}$, respectively, based on typical astrospheric values \citep{2000ApJ...537..304W,2005ApJS..159..118W}. We hold the astrospheric column density constant to avoid degeneracy with the free ISM column density. The astrospheric radial velocity is allowed to vary between $\pm$50 km s$^{-1}$, the same range as the ISM velocity. The astrospheric Doppler b parameter is allowed to vary between 20 and 30 km s$^{-1}$, based on the range of observed values.
\par
\begin{figure}[!ht]
\epsscale{1.15}
\plotone{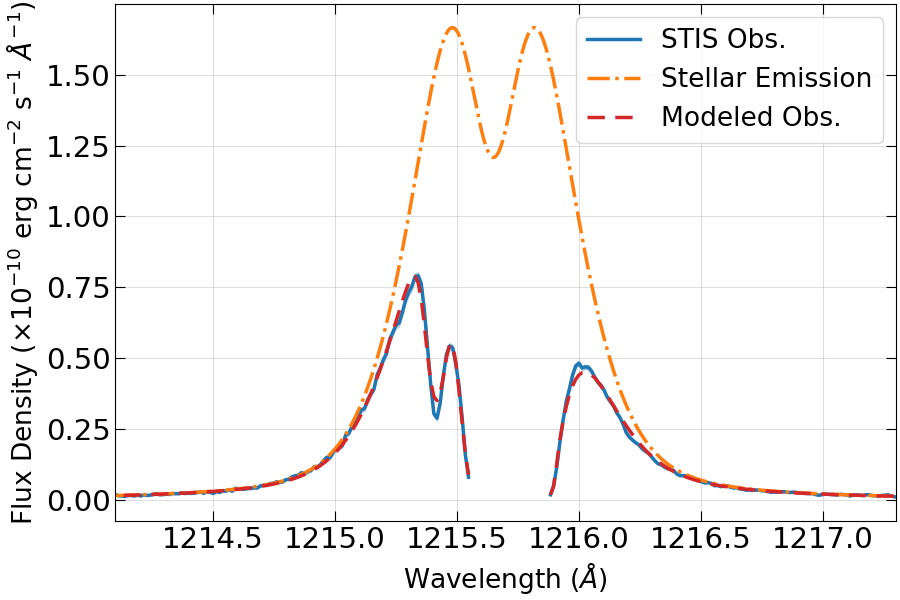}
\caption{The recovered Ly$\alpha$ profile of Procyon A from STIS/E140M data. Residual airglow located within the attenuated line core is excluded when determining the best fit to the data, and is not plotted here.}
\label{fig:LyA} 
\end{figure}

The best-fit result to the STIS E140M data is shown in Figure \ref{fig:LyA}, with best-fit parameters detailed in Table \ref{tab:params}. We find the best-fit stellar Ly$\alpha$ emission to be $(1.38\pm0.04) \times 10^{-10}$ erg cm$^{-2}$ s$^{-1}$, which is similar to $1.48 \times 10^{-10}$ erg cm$^{-2}$ s$^{-1}$ reported in \cite{2005ApJS..159..118W}. The best-fit column density is $18.059^{+0.014}_{-0.013}$, in excellent agreement with 18.06$\pm$0.01 calculated in \cite{1995ApJ...451..335L}. 

\subsection{FUV: SISTINE-2} \label{ss:sis2} 

The flux calibrated SISTINE-2 spectrum developed in Section \ref{sss:fluxcal} is used in the wavelength range 1029.0 - 1270.5 and 1299.6 - 1565.0 \AA, with the exception of the reconstructed Ly$\alpha$ profile described above. While Procyon A has been observed in the past with \textit{IUE}, \textit{FUSE}, and \textit{HST}/STIS at the same wavelengths as SISTINE-2, the data are taken at various epochs and do not individually capture the same bandpass as SISTINE-2. The \textit{FUSE} data was taken over a decade apart from the \textit{IUE} data, and another decade separates the STIS data from the \textit{FUSE} data. Stellar variability of F-type stars on the timescale of decades is currently poorly characterized for individual UV emission lines, so the simultaneous coverage of emission lines within the broad SISTINE-2 bandpass is preferred over the archival \textit{IUE}, \textit{FUSE}, and STIS spectra. We utilize \textit{FUSE} and STIS data only where SISTINE-2 does not have spectral coverage. 

\subsection{FUV: \textit{FUSE}} \label{ss:fuse} 

Below 1029 \AA, the effective area of SISTINE-2 becomes too small to reliably incorporate this region of the spectrum to the SED, and we instead utilize the \textit{FUSE} observation of Procyon A which covers 912 - 1029 \AA. Within this wavelength range lies the stellar \ion{H}{1} Ly$\beta$ emission line, and as with Ly$\alpha$, it is also obscured by the ISM. In addition to ISM contamination, geocoronal Ly$\beta$ emission also contaminates the profile, however this is mostly contained within the fully attenuated line core. 
\par
We modify the stellar reconstruction model developed in CA23 to function for \textit{FUSE} data, and include the astrospheric component described in Section \ref{sss:lya}. The Ly$\beta$ reconstruction is described below. While the remainder of the Lyman series lies within the \textit{FUSE} data, we only attempt a stellar reconstruction on Ly$\beta$ as we do not expect Ly$\gamma$ (972.5 \AA) or beyond to significantly contribute towards the total energy budget, and because a reconstruction is not possible on the fainter Lyman series lines \citep{2002ApJ...581..626R}. 

\subsubsection{Stellar \texorpdfstring{Ly$\beta$}{Lybeta} Reconstruction} \label{sss:lyb} 

We modify the Ly$\alpha$ reconstruction code described in Section \ref{sss:lya} to perform a reconstruction of the stellar Ly$\beta$ emission from the \textit{FUSE} data. For this reconstruction, we hold constant the best-fit results for the ISM and astrosphere found in the Ly$\alpha$ reconstruction, under the assumption that these properties do not vary significantly within the decade between the \textit{FUSE} and STIS observations. As a result, only the stellar emission parameters are varied when determining the best fit.
\par
We modify the self reversal parameters for the Ly$\beta$ fit to qualitatively represent the shape observed in solar Ly$\beta$ profiles \citep{1995ApJ...452..462M,2008A&A...490..307G,2009A&A...504..239T}, as was done for determining the self reversal parameters for Ly$\alpha$ in CA23. We adopt a self reversal depth parameter of 15\% and a width to be 30\% of the stellar emission's Gaussian FWHM.
\par
\begin{figure}[!ht]
\epsscale{1.15}
\plotone{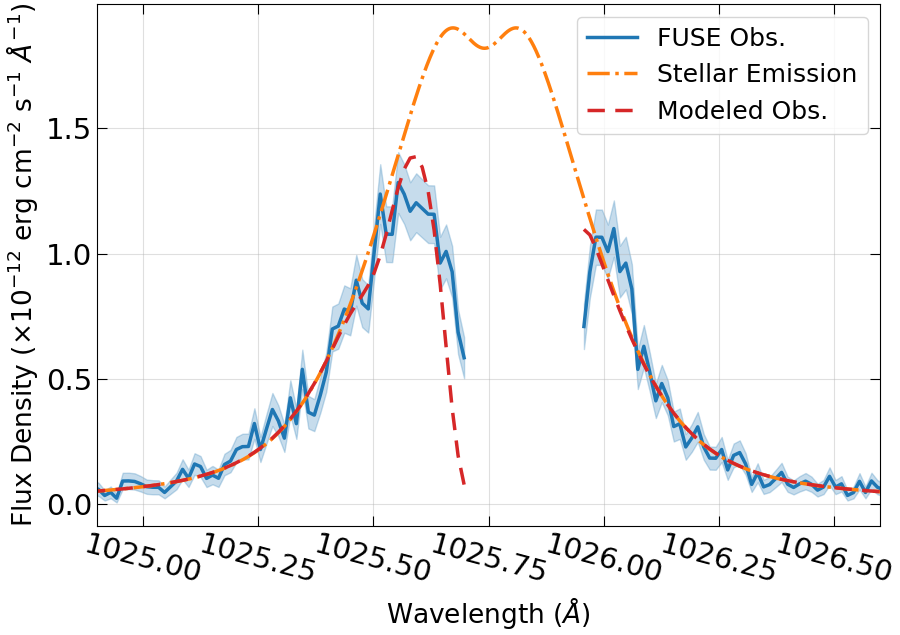}
\caption{The recovered Ly$\beta$ profile of Procyon A from \textit{FUSE} data. Unsubtracted airglow located within the attenuated line core is excluded when determining the best fit to the data, and is not plotted here.}
\label{fig:LyB} 
\end{figure}

As with the STIS E140M data, we exclude the central region of the emission line where ISM contamination is strongest. The narrow Ly$\beta$ airglow feature lies within this region, and is not included in the fit to the surrounding stellar data. The best-fit results are shown in Figure \ref{fig:LyB}, with the corresponding fit parameters in Table \ref{tab:params}. We find an integrated Ly$\beta$ flux of $(1.15^{+0.05}_{-0.04}) \times 10^{-12}$ erg cm$^{-2}$ s$^{-1}$. This corresponds to a stellar Ly$\alpha$/Ly$\beta$ ratio of $\sim$120, which is higher than the typical solar Ly$\alpha$/Ly$\beta$ ratio of $\sim$75, but within the observed range of solar ratios \citep{2012A&A...542L..25L}. The reconstructed Ly$\beta$ profile is used in the final SED in the wavelength range of 1024.75 - 1026.75 \AA.

\begin{deluxetable}{cccc}
\tablenum{2} 
\tablecaption{Best-fit parameters for the stellar reconstructions of Ly$\alpha$ and Ly$\beta$. \label{tab:params}}
\tablewidth{0pt}
\tablehead{
\colhead{Parameter} & \colhead{Ly$\alpha$} & \colhead{Ly$\beta$}
}
\startdata
Radial Velocity & \multirow{2}*{$-4.888^{+0.332}_{-0.340}$} & \multirow{2}*{$3.860^{+1.480}_{-1.471}$}  \\ 
(km s$^{-1}$) & & \\
Gaussian FWHM  & \multirow{2}*{$133.108^{+1.598}_{-1.575}$} & \multirow{2}*{$114.175^{+7.449}_{-7.807}$} \\ 
(km s$^{-1}$) & & \\
Lorentz FWHM & \multirow{2}*{$32.983^{+0.768}_{-0.784}$} & \multirow{2}*{$45.996^{+5.597}_{-4.538}$} \\ 
(km s$^{-1}$) & & \\
Flux Amplitude & \multirow{2}*{$-9.085^{+0.021}_{-0.020}$} & \multirow{2}*{$-11.295^{+0.047}_{-0.046}$} \\ 
(erg cm$^{-2}$ s$^{-1}$ \AA$^{-1}$) & & \\
Column Density & \multirow{2}*{$18.059^{+0.014}_{-0.013}$} & \multirow{2}*{-} \\ 
(10$^x$ cm$^{-2}$) & & \\
ISM Velocity & \multirow{2}*{$18.656^{+0.225}_{-0.256}$} & \multirow{2}*{-} \\ 
(km s$^{-1}$) & & \\
Astrosphere Velocity & \multirow{2}*{$3.386^{+3.501}_{-3.066}$} & \multirow{2}*{-} \\ 
(km s$^{-1}$) & & \\
Astrosphere b & \multirow{2}*{$23.198^{+2.329}_{-2.133}$} & \multirow{2}*{-} \\ 
(km s$^{-1}$) & & \\
Integrated Flux & \multirow{2}*{$1.38^{+0.04}_{-0.04} \times 10^{-10}$} & \multirow{2}*{$1.15^{+0.05}_{-0.04} \times 10^{-12}$} \\ 
(erg cm$^{-2}$ s$^{-1}$) & & \\
\enddata
\end{deluxetable}

\subsection{EUV: \texorpdfstring{Ly$\alpha$}{Lyalpha} Flux Relations} \label{ss:linsky} 

Below 912 \AA, photons ionize atomic hydrogen, resulting in significant obscuration of the stellar EUV spectrum by the ISM. Procyon A was observed with the \textit{Extreme Ultraviolet Explorer} (\textit{EUVE}), however the entire dataset could not be used in the SED; the stellar spectrum is heavily obscured by the ISM between 500 - 700 \AA\, and the instrument sensitivity is low longward of 300\AA. We instead use broadband proxies of stellar EUV emission within 500 - 912 \AA\ based on stellar and solar Ly$\alpha$ emission \citep{2014ApJ...780...61L}, as was done in the MUSCLES program \citep{2016ApJ...820...89F,2016ApJ...824..101Y}. Using the relations developed in \cite{2014ApJ...780...61L} with the reconstructed Ly$\alpha$ flux found in Section \ref{sss:lya}, four broadband EUV fluxes were calculated and integrated into the SED. 

\subsection{EUV: \textit{EUVE}} \label{ss:euve} 

We utilize the remainder of the \textit{EUVE} observation of Procyon A in the wavelength range 90 - 500 \AA\ \citep{1995ApJ...443..393D}. This portion of the EUV spectrum still suffers from ISM obscuration, however the correction factors are relatively small and a reliable reconstruction is possible. We utilize the best-fit column density from the Ly$\alpha$ reconstruction to derive the attenuation profile of the ISM in the line of sight of Procyon A. The EUV spectrum is then corrected for ISM attenuation, and is merged into the SED.
\par
We use the best-fit hydrogen column density from the stellar Ly$\alpha$ reconstruction to estimate the \ion{He}{1} and \ion{He}{2} column densities. From \cite{1995ApJ...455..574D}, we adopt a column density ratio N$_{H I}$/N$_{He I}\approx$14. \cite{1995ApJ...455..574D} additionally provides upper limits to the \ion{He}{2} ionization fraction along several lines of sight, and we assume an ionization fraction of 70\%. This allows us to calculate a \ion{He}{2} column density from the \ion{He}{1} column density, which in turn is estimated from the \ion{H}{1} column density.
\par
We calculate the hydrogenic photoionization cross sections using:

\begin{equation}
\sigma_\nu=\frac{7.91\times10^{-18}}{Z^2}\left(\frac{\nu_1}{\nu}\right)^3g_{1f}
\label{eqn:sigma}
\end{equation}

\noindent where $Z$ is the atomic number, $\nu_1$ is the ionization frequency, $\nu$ is the photon frequency, $g_{1f}$ is the bound-free gaunt factor from the ground state, and $\sigma_\nu$ is in cm$^2$. We use Equation 5-7 in \cite{1978ppim.book.....S} to calculate the bound-free gaunt factors:

\begin{equation}
g_{1f}=8\pi\sqrt{3}\frac{\nu_1}{\nu}\frac{e^{-4zcot^{-1}z}}{1-e^{-2\pi z}}
\label{eqn:gaunt}
\end{equation}

\noindent where

\begin{equation}
z^2=\frac{\nu_1}{\nu-\nu_1}
\label{eqn:z}
\end{equation}

\noindent We use the \ion{He}{1} photoionization cross section from \cite{1966AdAMP...2..177S}. Using the column density relations and the photoionization cross sections for \ion{H}{1}, \ion{He}{1}, and \ion{He}{2}, we estimate the transmission profile across the \textit{EUVE} spectrum. The stellar signal is corrected for ISM attenuation by dividing the \textit{EUVE} spectrum with the transmission profile.

\subsection{X-Ray: Chandra} \label{ss:xray} 

Procyon A has been observed in the X-ray regime using the Low Energy Transmission Grating (LETG) and High Resolution Camera, optimized for Spectroscopy (HRC-S) on board the Chandra X-ray Observatory. We include Chandra data from 10 - 90 \AA\ in the SED. The downloadable Chandra data products are given in photon counts per pixel for both the science and background spectra, and thus required flux calibration for integration with the SED. 
\par
A wavelength solution can be generated from the data available in the Chandra fits file. This along with the recorded exposure time is applied to the science and background spectra to obtain units of erg s$^{-1}$ \AA$^{-1}$. We utilize the CIAO software package to generate effective area curves, and a background subtracted and flux calibrated X-ray spectrum of Procyon A is produced in erg cm$^{-2}$ s$^{-1}$ \AA$^{-1}$. We compare the integrated fluxes of emission features within our flux calibrated X-ray spectrum to those in \cite{2002A&A...389..228R}, and find good agreement. This final component is integrated to the SED of Procyon A, and we see good agreement at the interface between the Chandra and \textit{EUVE} data. 

\section{Terrestrial Planets around \\ F-Type Stars} \label{s:exoplanets} 

With the development of a full SED of the mid F-type star Procyon A, several avenues are now open. The SED can be used as a template for other aging mid F-type stars in exoplanet atmosphere models, EUV reconstructions, modeling spectral evolution of dwarf stars, and all other studies which require the stellar input of such a star. In this section, we will take a brief inventory of currently known exoplanets around F-type stars (Section \ref{ss:known}), and then explore a model for detecting the hydrogen exosphere of an Earth-like exoplanet in the habitable zone (HZ) of mid-F stars, using Procyon A as a template (Sections \ref{ss:earth} and \ref{ss:detect}). 

\subsection{Known Planets around F-Type Stars} \label{ss:known} 

As of writing, there are currently 191 confirmed exoplanets orbiting around F-type stars \citep{NEA}. There is a wide diversity among these planets, consisting of Super-Earths (Kepler-21 b, HD 106315 b, HD 28109 b, and TOI-411 b), sub-Neptunes (e.g. TOI-1670 b, HD 5278 b), Neptune-like planets (e.g. K2-98 b, HD 106315 c), Hot Giants (e.g. WASP-28 b, KELT-6 b), Warm Giants (e.g. TOI 1670 c, HD 332231 b) and Cold Giants (e.g. HD 148164 b, c). These planets are found as close as $\sim$4 R$_{\odot}$ to their host stars and as far out as 650 AU. These planetary systems have been found as nearby as 13.4 pc and as distant as 2.35 kpc. 
\par
We use a HZ calculator\footnote{\url{https://personal.ems.psu.edu/~ruk15/planets/}} to determine conservative and optimistic HZ distances around Procyon A \citep{2013ApJ...765..131K,2013ApJ...770...82K,2014ApJ...787L..29K}. For this calculation, we adopt the values of T$_{eff}$ = 6530 K and L$_*$ = 6.93 L$_{\odot}$ from \cite{2004A&A...413..251K}. We find the conservative HZ  of Procyon A to be bounded by 2.391 - 4.152 AU, while the optimistic HZ is bounded by 1.888 - 4.375 AU. 
\par
We use representative values for main-sequence F-type stars in the HZ calculator, and find that the HZs of F-type stars generally range roughly between 1 - 4 AU \citep{2013ApJS..208....9P}.\footnote{\url{http://www.pas.rochester.edu/~emamajek/EEM_dwarf_UBVIJHK_colors_Teff.txt}} Within this range, there are 26 confirmed exoplanets orbiting within the HZs of their F-type host stars. The masses and radii of these planets are generally not known, apart from $\upsilon$ Andromedae d (10.25 M$_J$) and HD 206893 c (1.46 R$_J$, 12.7 M$_J$). The other known HZ planets have $m_p$ $sin$ $i$ values from radial velocity measurements, and all are roughly Jupiter mass or larger. The four known Super-Earths around F dwarfs all orbit their host star within the orbit of Mercury. The future \textit{HWO} will prioritize the observations of F-, G-, and K-type stars, the best candidates for hosting Earth-like exoplanets, and is expected to discover on the order of 10 - 100 exoplanets orbiting around F dwarfs in currently unexplored parameter spaces \citep{2019arXiv191206219T,2020arXiv200106683G}.

\subsection{An Earth-like Planet Around Procyon A} \label{ss:earth} 

\begin{figure}[!ht]
\epsscale{1.15}
\plotone{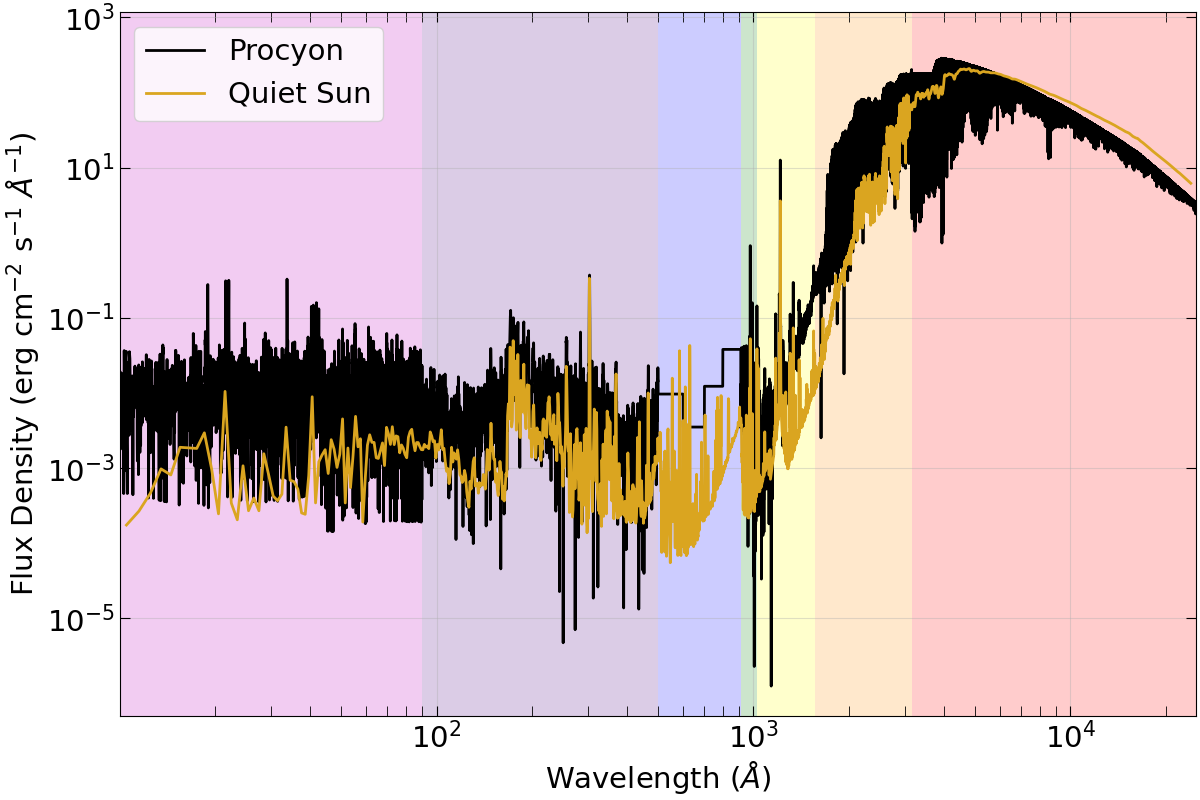}
\caption{The SED at 2.632 AU from Procyon A, where a hypothetical Earth-like exoplanet would receive the same insolation as Earth does from the Sun. The SED of the quiet Sun as observed from Earth is shown for comparison \citep{2009GeoRL..36.1101W}. Colored regions are the same as in Figure \ref{fig:SED}.}
\label{fig:SED_1AU} 
\end{figure}

While there are currently no known habitable planet candidates around F-type stars, including Procyon A, we proceed with a hypothetical Earth-like exoplanet orbiting within the HZ of Procyon A. We calculate the orbital distance from Procyon A where this hypothetical planet would receive the same insolation as the Earth as 2.632 AU \citep{2013ApJ...765..131K,2013ApJ...770...82K}. We scale the SED to this distance, and a comparison to the quiet Sun can be seen in Figure \ref{fig:SED_1AU} \citep{2009GeoRL..36.1101W}. We note the enhancement of XUV emission of Procyon A compared to the Sun. At solar maximum the Sun displays an order of magnitude enhancement in XUV emission \citep{Tom_Woods2020-pj}, similar to the observed XUV emission of Procyon A. We explore the impact of the XUV enhancement in this section.  

\subsubsection{Atmospheric Modeling} \label{sss:model} 

We develop a simple model of the atmosphere of an Earth-like exoplanet and its response to different levels of XUV emission from its host star. We primarily focus on the thermosphere and exosphere of this planet; the thermosphere of the Earth is heated by solar XUV radiation, expanding and contracting in response to solar activity. The exobase is the boundary where the thermosphere ends and the exosphere begins, and is commonly defined as the altitude above the Earth at which the local scale height is comparable to the mean free path of particles in the atmosphere. Atmospheric temperature in the exosphere is nearly constant. Particles in the exosphere are effectively collisionless, and will either fall back to the exobase or will escape the atmosphere via Jeans escape. The exobase altitude changes as a response to thermospheric heating from incoming XUV radiation.   
\par
To model the thermosphere and exosphere of this hypothetical planet, we use the Jacchia 1977 (J77) model of the Earth's atmosphere \citep{1977SAOSR.375.....J,1983CeMec..29....3D}. The only model parameter is the temperature of the exosphere, which can be determined by the level of XUV radiation received by the planet's host star \citep{2022ApJ...937...72N}. \cite{2017GeoRL..4411706K} shows observational evidence that the \ion{H}{1} exosphere of the Earth extends to $>$ 38 R$_{\oplus}$. We calculate the J77 atmosphere model up to 50 R$_{\oplus}$, however we force the number density of \ion{H}{1} to decrease to 1 cm$^{-3}$ between 38 - 50 R$_{\oplus}$ via exponential decay. This allows for a smooth transition with the lower region of the model atmosphere. 
\par
Before applying the model described above to Procyon A, we first test it with the Earth-Sun system. We assume an exosphere temperature of 900 K, representing average solar activity. We calculate the altitude of the exobase of this model to be 325 km above the surface of the Earth. At the exobase, we calculate the Jeans escape parameter as 

\begin{equation}
\lambda_J=\frac{GMm}{k_BT_{exo}r_{exo}}
\label{eqn:jPara}
\end{equation}

\noindent where G is the gravitational constant, M is the mass of the Earth, m is the mass of \ion{H}{1}, k$_B$ is the Boltzmann constant, and T$_{exo}$ and r$_{exo}$ are the temperature and altitude of the exobase, respectively. We also calculate the Jeans escape flux at the exobase

\begin{equation}
\Phi_{J}=\frac{n_{exo,HI}v_s}{2\sqrt{\pi}}(1+\lambda_J)e^{-\lambda_J}
\label{eqn:jFlux}
\end{equation}

\noindent where $n_{exo,HI}$ is the \ion{H}{1} number density at the exobase and v$_s$ is the most probable speed of \ion{H}{1} atoms at the exobase. 
\par
At the exobase, we find $\lambda_J$ = 8.14 and $\Phi_J$ = 6.01 $\times 10^7$ cm$^{-2}$ s$^{-1}$ (568 g s$^{-1}$). Under these conditions, the Earth does not experience atmospheric blowoff. The calculated exobase altitude in our model is smaller than the 500 km that is typically quoted for the Earth \citep[e.g.][]{1973JGR....78.1115V,2016NatCo...713655Q,2017GeoRL..4411706K}, however the atmospheric properties at this altitude are within the range of observed values at the Earth's exobase \citep{1973JGR....78.1115V,1978GeoRL...5...29V,2017GeoRL..4411706K}. 
\par
We integrate the SED developed in Section \ref{s:SED} between 10 - 1180 \AA\ and find the XUV emission of Procyon A to be 12.61 $\pm$ 0.08 mW m$^{-2}$. Assuming solar XUV emission to be 5.1 mW m$^{-2}$ \citep{2008JGRE..113.5008T}, the XUV emission of Procyon A is (2.47 $\pm$ 0.02) $\times$ solar XUV emission. If instead we use 6.6 mW m$^{-2}$ \citep{2020AJ....160..237F}, we get (1.91 $\pm$ 0.01) $\times$ solar XUV emission.
\par
Within this range of possible XUV emission values, we determine that exobase temperatures on a hypothetical Earth-like planet around Procyon A would range from 1700 - 2300 K \citep{2022ApJ...937...72N}. We run our atmosphere model within this range of exosphere temperatures, with a grid spacing of 100 K. The results of this model grid are given in Table \ref{tab:exosphere}. For context, the exosphere temperature on Earth during solar maximum is $\sim$1600 K \citep{1991JGR....96.1159H}.

\begin{deluxetable}{cccccc}
\tablenum{3} 
\tablecaption{Atmosphere model results of the hypothetical Earth-like exoplanet orbiting in the HZ of Procyon A for various exosphere temperatures. \label{tab:exosphere}}
\tablewidth{0pt}
\tablehead{
\colhead{T$_{exo}$} & \colhead{r$_{exo}$} & \colhead{n$_{exo,HI}$} & \colhead{$\lambda_{J}$} & \colhead{$\Phi_{J}$} & \colhead{$|\Delta$F/F$|_{max}$}
\\
\colhead{(K)} & \colhead{(km)} & \colhead{($\times$10$^{4}$ cm$^{-3}$)} & \colhead{} & \colhead{($\times$10$^{3}$ g s$^{-1}$)} & \colhead{(\%)}
}
\startdata
1700 & 517 & 2.72 & 4.15 & 3.29 & 2.84  \\
1800 & 543 & 2.30 & 3.90 & 3.52 & 2.87  \\
1900 & 568 & 1.96 & 3.68 & 3.71 & 2.88  \\
2000 & 595 & 1.69 & 3.48 & 3.86 & 2.89  \\
2100 & 621 & 1.47 & 3.30 & 3.99 & 2.90  \\
2200 & 648 & 1.29 & 3.14 & 4.08 & 2.90  \\
2300 & 676 & 1.14 & 2.99 & 4.15 & 2.89  \\ \hline
900 & 325 & 20.9 & 8.14 & 0.57 & 6.18  \\
\enddata
\tablecomments{The final row shows Earth around the Sun for comparison.}
\end{deluxetable}

The model results show that as XUV radiation (and therefore exobase temperature and altitude) increases, the number density at the exobase decreases. Despite this, the Jeans escape flux continues to increase with increasing exobase temperature. This feature arises due to the exobase occurring at higher altitudes with increasing XUV radiation, which increases the surface area of escape at a rate faster than the number density  of the escaping \ion{H}{1} decreases. The Jeans escape parameter decreases with hotter, higher altitude exospheres, however none of the models we considered reach the critical value below 1.5, where atmospheric blowoff begins \citep{2008JGRE..113.5008T}. 
\par
\begin{figure}[!ht]
\epsscale{1.15}
\plotone{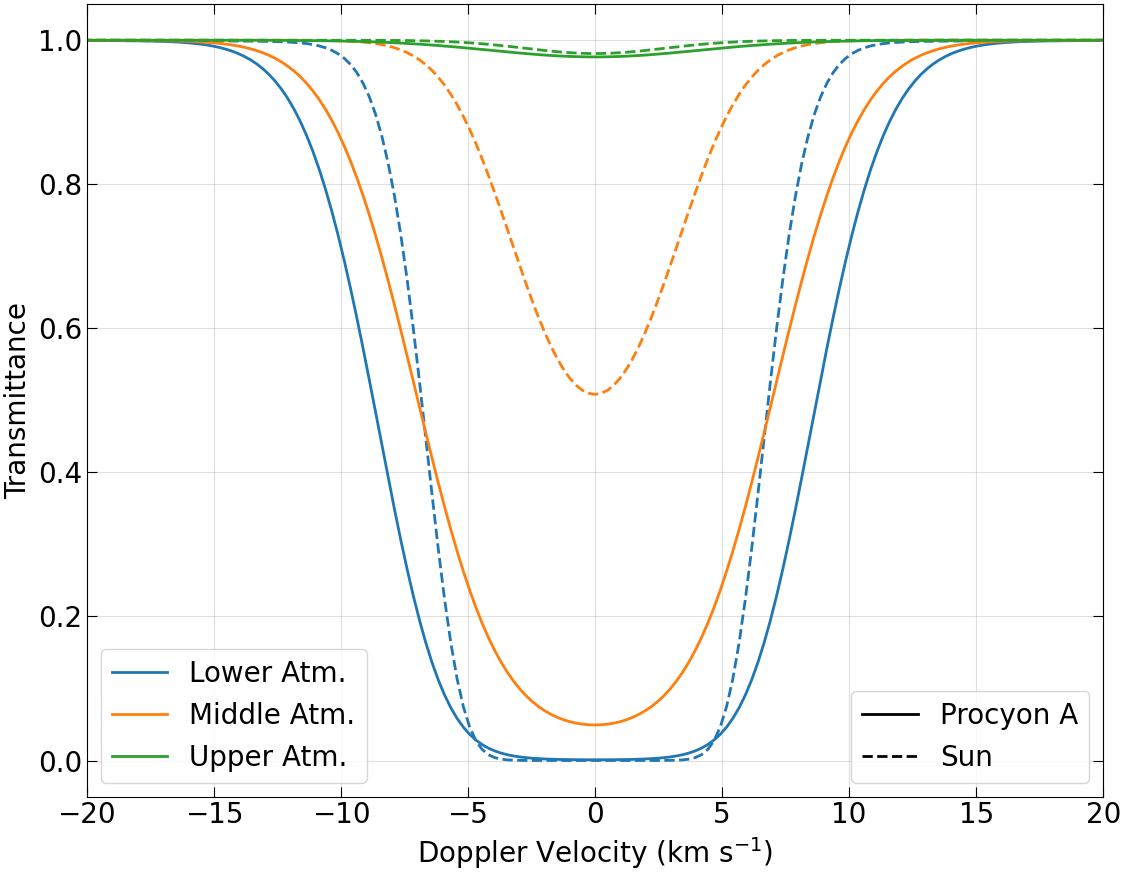}
\caption{Transmission curves in the lower (blue), middle (orange) and upper (green) atmospheres of an Earth-like planet around Procyon A (solid) and the Earth around the Sun (dashed). Like-colored curves are calculated at the same altitude. The enhanced XUV flux of Procyon A results in generally broader and deeper transmission features throughout the atmosphere.}
\label{fig:Transmission} 
\end{figure}

We calculate the photoabsorption cross section of both \ion{H}{1} and \ion{D}{1} to stellar Ly$\alpha$ emission as a function of both wavelength and local atmospheric temperature. We calculate the \ion{D}{1} number density assuming the bulk Earth D/H ratio of 1.49$\times 10^{-4}$ \citep{2017RSPTA.37550390H}. We then calculate the optical depth $\tau$ along several lines of sight throughout the atmosphere, and combine these to construct atmospheric transmission curves, with respect to stellar Ly$\alpha$ photons, at various altitudes. Example transmission curves along lines of sight at different altitudes are shown in Figure \ref{fig:Transmission} for the Sun-Earth and Procyon A-Earth cases. The observed transit depth ($\Delta$ F/F) from the atmosphere passing in front of its host star is then calculated as 

\begin{equation}
\frac{\Delta F}{F} = \frac{-\int 2b(1-e^{-\tau})db}{R_*^2 - R_p^2}
\label{eqn:transit_depth}
\end{equation}

\noindent where $\tau$ is the wavelength dependant optical depth along a given line of sight, making its closest approach b (in km) to the planet. Integrating over the lines of sight as a function of b results in the fractional change of starlight due to obscuration of the hydrogen exosphere. The stellar radius is R$_*$ and the planetary radius is R$_p$, both in km.

\begin{figure}[!ht]
\epsscale{1.15}
\plotone{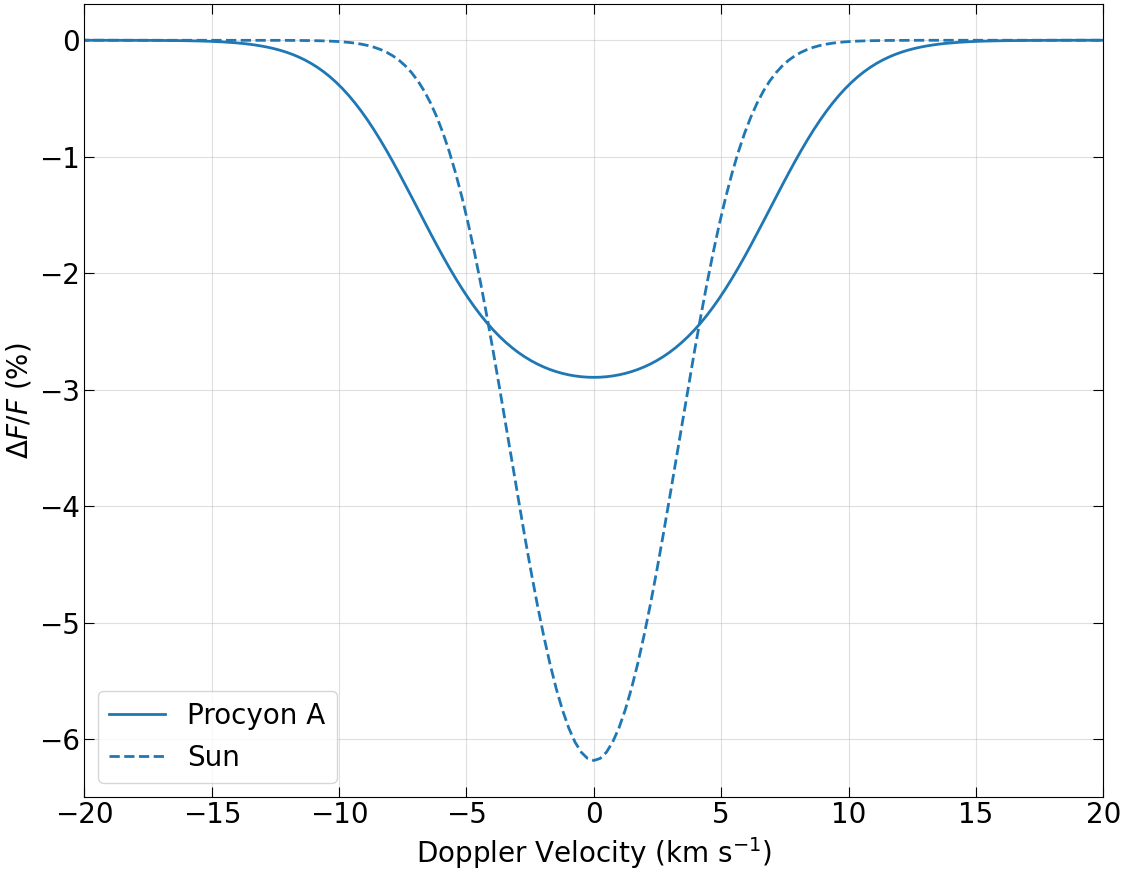}
\caption{The transit depth observed from an Earth-like exoplanet passing in front of Procyon A (solid) and the Earth passing in front of the Sun (dashed). The larger size of Procyon A suppresses the depth of the transit signal.}
\label{fig:DF_F} 
\end{figure}

Across all models, the transit depth is approximately constant. We note that the transit depth of the Earth passing in front of the Sun is deeper than when the planet passes in front of Procyon A. While the number density of \ion{H}{1} at a given altitude is generally larger in the thermosphere and exosphere as XUV emission increases, the transit depth decreases in this case due to the larger radius of Procyon A (2.048 $\pm$ 0.025 R$_{\odot}$, \citealt{2004A&A...413..251K}). If the Sun were to irradiate the Earth with the amount of XUV radiation delivered by Procyon A (e.g. for a younger or more active solar-type star), a $\sim$12\% transit depth would be observed. See Figure \ref{fig:DF_F} for a comparison between the transit depths of the Earth transiting the Sun and the Earth-like exoplanet transiting Procyon A. 

\subsubsection{A Young Procyon A} \label{ss:youngProc} 
Younger, more active solar-type stars have been observed to emit one to two orders of magnitude more flux than the Sun currently emits across the XUV bandpass \citep[See][]{2005ApJ...622..680R,2018ApJS..239...16F}. The youngest, most active main sequence F, G, K, and M dwarfs reach a saturation point where flux no longer increases with decreasing age \citep[See][where stellar rotation period acts as a proxy for stellar age]{2011ApJ...743...48W,2018ApJS..239...16F,airglow}. The calculation of enhanced XUV emission from a Young Procyon A, as well as speculation on how this impacts the atmosphere of an Earth-like planet in the HZ, is briefly explored here.
\par
To predict the XUV output of a Young Procyon A, we first calculate the integrated flux within the 90 - 360 \AA\ bandpass. Figure 6 of \cite{2018ApJS..239...16F} displays the relationship between stellar rotation period and F$_{90-360}$/F$_bol$, the ratio between EUV flux in the 90 - 360 \AA\ bandpass and the bolometric flux. With a rotation period of 23 $\pm$ 2 days \citep{1991ApJ...375..704A} and a measured F$_{90-360}$/F$_bol$ of (2.37 $\pm$ 0.08) $\times$ 10$^{-6}$, the location of Procyon A within this parameter space is consistent with that of other F-type and G-type stars. We estimate the saturated F$_{90-360}$/F$_bol$ value within this bandpass by taking the average flux ratio values of the seven stars within the saturated regime in Figure 6 of \cite{2018ApJS..239...16F}, found to be 2.36 $\times$ 10$^{-4}$. We find that a Young Procyon A would have $\sim$100$\times$ the flux within this bandpass, and we assume that this scale factor roughly holds throughout the remaining XUV bandpass based on observed trends of solar-type stars of different activity levels \citep{2005ApJ...622..680R}.
\par
Properly modeling the impact that a two order of magnitude XUV enhancement would have on a terrestrial atmosphere is beyond the scope of our simple model presented here. It is probable that the atmosphere will enter the hydrodynamic escape regime and evaporate. Just as with Earth and Venus, and as is speculated for terrestrial planets around M dwarfs \citep{2020PNAS..11718264K,2020MNRAS.496.3786M,2020AJ....160..237F}, the development of a secondary atmosphere is a possibility for planets orbiting around F-type stars. For the remainder of this work, we will assume that these terrestrial exoplanets have developed secondary, Earth-like atmospheres.

\subsection{Detectability with Space Telescopes} \label{ss:detect} 

We now assess the detectability of the atmospheres of terrestrial planets orbiting around mid F-type stars with current and future FUV spectrographs. We consider the COS and STIS instruments on board the \textit{Hubble Space Telescope} (\textit{HST}), the two currently operational FUV spectrographs. As a representative for an FUV spectrograph on board the future \textit{HWO}, we consider the \textit{LUVOIR} Ultraviolet Multi-Object Spectrograph (LUMOS) instrument \citep{2017SPIE10397E..13F} developed for the Architecture B of the \textit{Large Ultraviolet / Optical / Infrared Surveyor} (\textit{LUVOIR}) mission concept \citep{2019arXiv191206219T}.
\par
We multiply the reconstructed Ly$\alpha$ profile, developed in Section \ref{sss:lya} for the SED, with the transmission curves to model the in-transit spectrum. We note that for the ISM attenuation present along the line of sight towards Procyon A, the entire transit signal would be completely attenuated by the ISM. Because of this, the only stellar systems for which this signal could be detected are stars with radial velocities (RVs) greater than 80 km s$^{-1}$, where the transit signal can be separated from the ISM attenuation \citep{2022ApJ...926..129Y}.
\par
We use the reconstructed stellar emission as our out of transit signal, and apply the T$_{exo}$ = 1700 K transmission curve to the stellar emission as our in-transit signal. This transmission curve was chosen because it will produce the weakest transit signal. The in and out of transit Ly$\alpha$ profiles are converted from erg cm$^{-2}$ s$^{-1}$ \AA$^{-1}$ to counts s$^{-1}$ pixel$^{-1}$ by using the effective areas and dispersion relations for three FUV gratings: COS/G130M \citep{2022cosi.book...14J}, STIS/E140M \citep{2022stii.book...21P}, and LUMOS/G120M \citep{2019arXiv191206219T}. For COS/G130M and STIS/E140M, we convolve the stellar emission with their measured LSFs.\footnote{\url{https://www.stsci.edu/hst/instrumentation/cos/performance/spectral-resolution}}\textsuperscript{,}\footnote{\url{https://www.stsci.edu/hst/instrumentation/stis/performance/spectral-resolution}} For LUMOS/G120M, we construct a Gaussian LSF based on the reported resolving power for this grating mode \citep{2019arXiv191206219T}.
\par
For COS/G130M and STIS/E140M, we implement photon noise and sources of background noise provided in their respective instrument handbooks \citep{2022cosi.book...14J,2022stii.book...21P}, however airglow is only considered for the COS/G130M observations. The narrow STIS slit makes an alignment of the high RV stellar profile and the airglow contamination unlikely, however the broad COS aperture makes airglow contamination for high RV stars plausible. We consider high, low, and zero airglow contamination cases for the COS/G130M observations. Apart from cases where airglow is considered, photon noise dominates the error terms in each pixel. For the LUMOS/G120M observations, we only consider photon noise as background terms for the LUMOS instrument are not currently characterized, however they are expected to be negligible as with COS/G130M and STIS/E140M.
\par
We simulate FUV transit observations for the three spectrographs at various distances to the target star systems. For the \textit{HST} instruments, we consider stellar systems that are the current distance to Procyon A at 3.51 pc \citep{2007A&A...474..653V}, and systems 10 pc and 25 pc away. For STIS/E140M, we follow a 5 orbits per visit observing strategy as is required to protect and prolong the health the STIS FUV detector.\footnote{\url{https://hst-docs.stsci.edu/hsp}} For LUMOS, we consider distances within $\sim$100 pc (the local ISM). 
\par
Unlike previous FUV transit observations, which are of planets close to their host stars, the orbital period of this hypothetical planet is 3.5 years and would have a transit time of 35.5 hours (127.8 ks) as calculated using the technique described in \cite{2010MNRAS.407..301K}, requiring multiple days of nearly continuous observation. Continuous \textit{HST} observations are not possible due to the limited amount of time allotted for observations per \textit{HST} orbit. We account for this in our light curve model and only allow a total of 2400 s of observing time per orbit. To protect the STIS FUV detectors from overillumination of the bright stellar Ly$\alpha$ emission, we separate each \textit{HST} visit with four ``dead orbits'' at the cost of further reductions in observing time. For COS and LUMOS, we do not consider these constraints, and assume that near-continuous observation before, during, and after the transit is possible.

\subsubsection{COS/G130M Observations} \label{sss:cos} 

While the Ly$\alpha$ profile of a high RV star will no longer suffer from ISM attenuation, the emission will be so strong that the Ly$\alpha$ line alone will exceed the global count limit of the COS detector for targets 3.51 and 10 pc away. Therefore we only consider the 25 pc case for COS/G130M observations, where the count rate is low enough to be safely observed. The highest achievable SNR with COS/G130M is $\sim$47.6 \citep{2022cosi.book...14J}. This is achieved in a 100 s exposure, however this is not enough to resolve the shallow and narrow transmission signal of the planet's atmosphere to a 3$\sigma$ significance. Multiple 100 s observations must be coadded together to separate the in and out of transit signals.
\par
We find that the minimum observing time required to make a 3$\sigma$ transit detection with COS/G130M to be 26.4 ks, in the case where minimal airglow contamination is present. In this case, the transit signal is marginally detectable, where only two coadded exposures of sufficient SNR can be taken within the transit. A high level of airglow contamination would require 480 ks minimum of observing time, and is unfeasible within a single orbit, requiring several days of dedicated observing time every 3.5 years to build up the transit signal to a 3$\sigma$ significance. The detection of such a planetary atmosphere, even in the best case is difficult with COS/G130M under the considered restraints.

\subsubsection{STIS/E140M Observations} \label{sss:stis} 

\begin{figure*}[!ht]
\epsscale{1.05}
\plotone{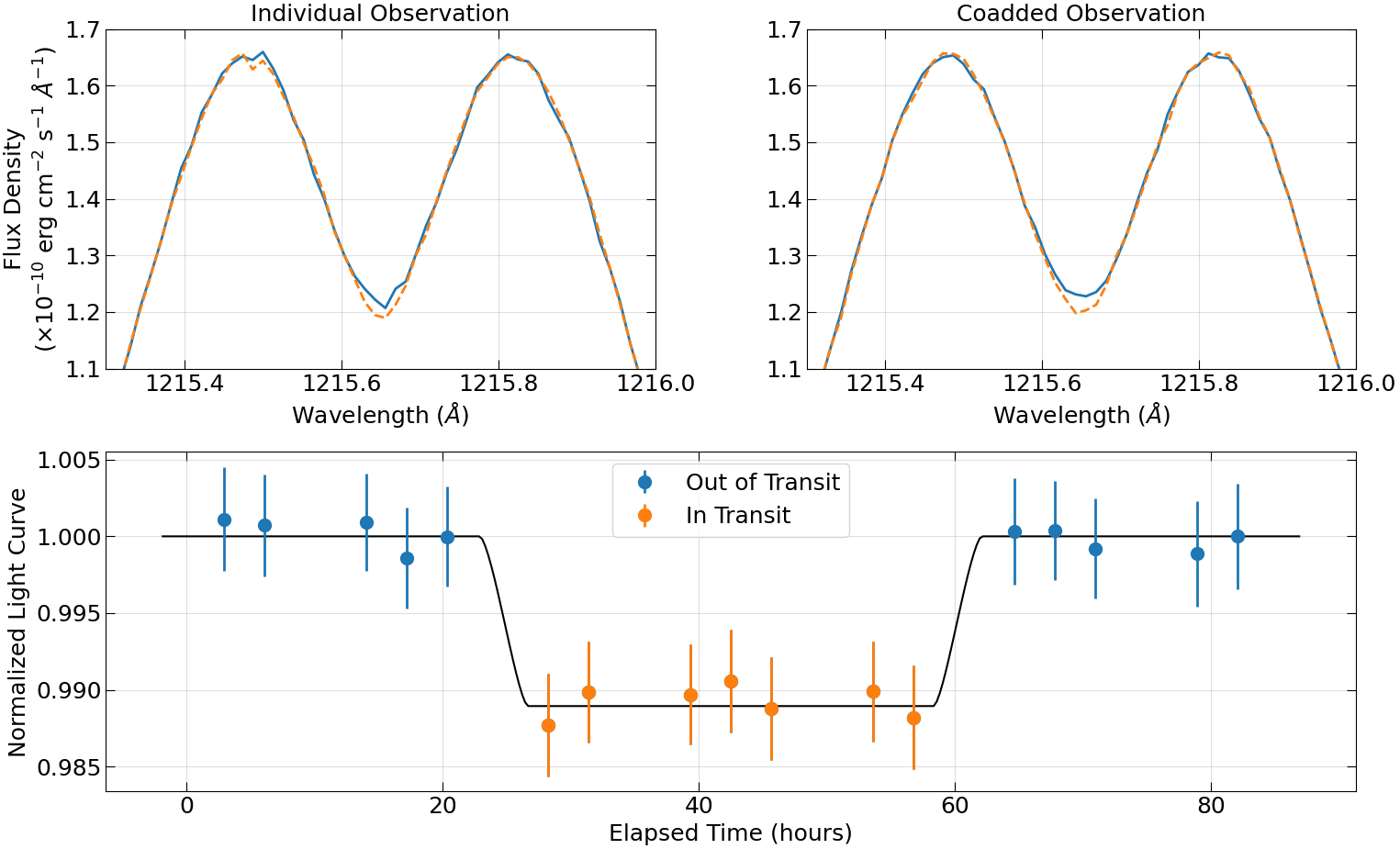}
\caption{A simulated STIS/E140M observation of an Earth-like exoplanet transiting a high RV F-type host star 3.51 pc away. The upper left panel shows the out of transit (blue) and in transit (orange) Ly$\alpha$ profiles from a single 2400 s observation. The upper right panel shows coadded observations totaling 4800 s. The coadded spectra are used to simulate the light curve in the bottom panel, where 3$\sigma$ error bars are drawn on each point.\vspace{0.5cm}}
\label{fig:LC_STIS} 
\plotone{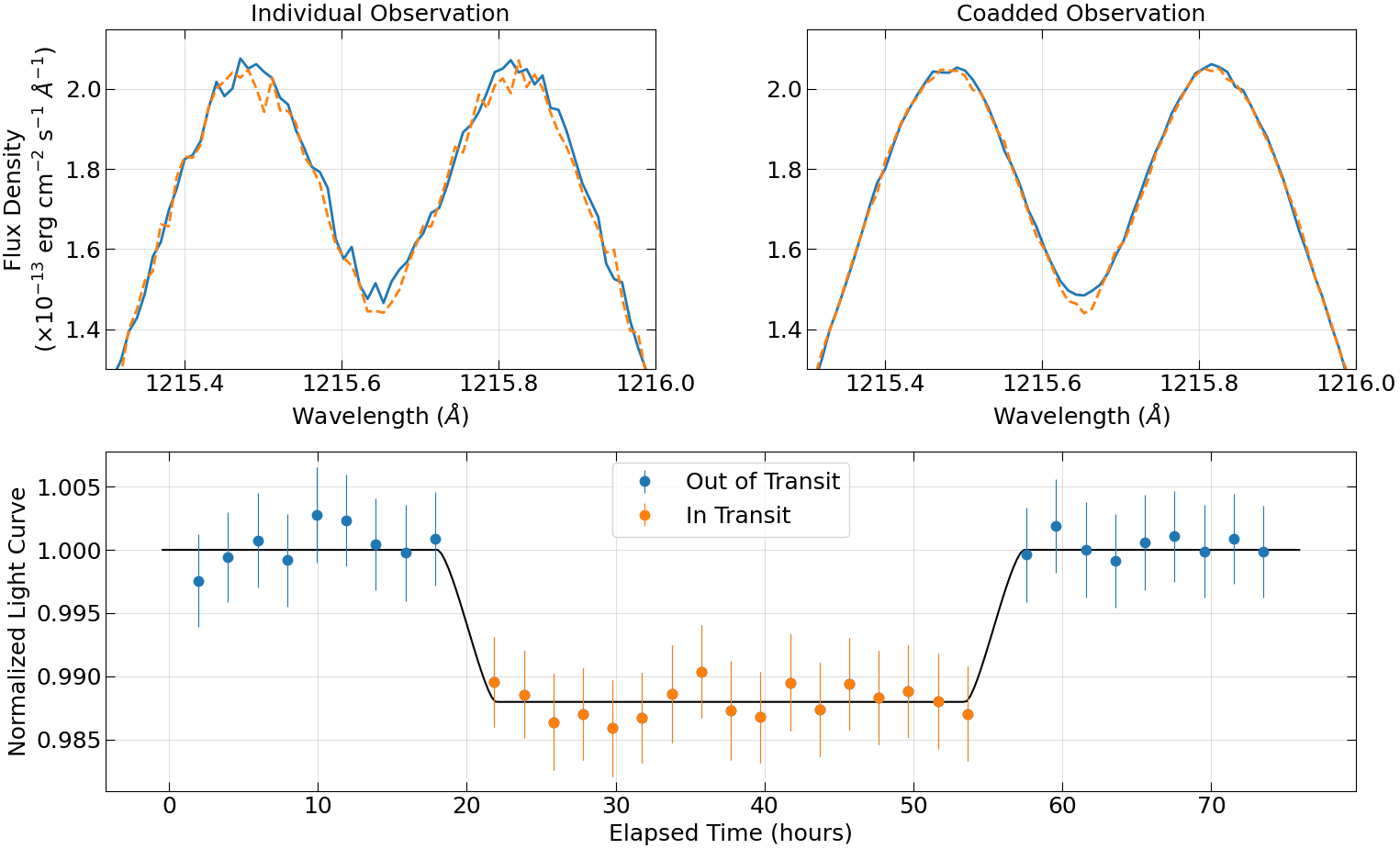}
\caption{A simulated observation with panels as described in Figure \ref{fig:LC_STIS}, but for LUMOS/G120M. The target is located 100 pc away. Individual observations are 650 s long, and the coadded observation totals to 7150 s. Error bars represent 3$\sigma$ errors.}
\label{fig:LC_LUMOS} 
\end{figure*}

The reduced effective area of the STIS/E140M grating mode eliminates global count rate issues for the 3.51 and 10 pc cases, and airglow contamination is no longer a consideration. Similar to COS/G130M, we find that multiple transits are required for the 10 and 25 pc cases, primarily due to count rates being too low to achieve the necessary SNR for a 3$\sigma$ detection within the observational constraints. For the 3.51 pc case, a 3$\sigma$ detection is possible within a single transit. An example of such a detection is shown in Figure \ref{fig:LC_STIS}. For simplicity, we only simulate points completely in or completely out of transit, and do not attempt limb darkening corrections. The \texttt{batman} python package is used to illustrate what the underlying transit light curve might look like \citep{2015PASP..127.1161K}. 
\par
A time series consisting of individual 2400 s exposures taken within one \textit{HST} orbit is unable resolve the planetary transit. To achieve a 3$\sigma$ detection between in and out of transit signals, a time series where each point is a coaddition of two successive exposures is required. While we do consider that there is a SNR limit for each 2400 s exposure, the limiting SNR for STIS/E140M is much higher than COS/G130M (SNR = 390 at best, \citealt{2022stii.book...21P}) and is not a constraint in this case.

\subsubsection{LUMOS/G120M Observations} \label{sss:lumos} 

We use the LUMOS-B instrument design described in \cite{2019arXiv191206219T} to represent an FUV spectrograph on board \textit{HWO}. The effective area of LUMOS/G120M is $\sim$40$\times$ greater than COS/G130M and $\sim$640$\times$ greater than STIS/E140M at Ly$\alpha$ wavelengths, providing a significant improvement in the detectability of terrestrial exoplanet atmospheres. While LUMOS/G120M is capable of providing near continuous observations of the transit, there is a limit to the achievable SNR in a single exposure with its MCP detectors. We estimate this limiting SNR to be 100 for the LUMOS detector based on the theoretical performance of MCP detectors \citep{2010hstc.workE...4A}, and from this a maximum exposure time is calculated for individual exposures. We find that transit signals can be measured at several $\sigma$ significance levels for all three distances considered for \textit{HST} due to the large effective area provided by the \textit{LUVOIR}-B architecture, with the caveat that bright object limits have not yet been established for LUMOS and it is possible that these stars will produce local rate limit violations on the \textit{HWO} detector(s). With this being said, several possible observing strategies can be implemented due to the additional flexibility compared with the \textit{HST} observational constraints. 
\par
We consider a high RV F-type star at a distance of 100 pc, the outer limit of the local ISM. Beyond this distance, the large column densities along most lines of sight will fully attenuate the stellar Ly$\alpha$ line even for large radial velocities. For a target 100 pc away, the limiting SNR is achieved in a 650 s exposure. Creating 7150 s coadditions would allow for a 3$\sigma$ detection of the planetary transit. The light curve generated by this observing stratedy is presented in Figure \ref{fig:LC_LUMOS}. The light curve is more detailed than the STIS/E140M detection for a target four times as distant, demonstrating the capabilities of a $\sim$6 m space telescope as recommended by the Astro2020 Decadal Survey.

\subsubsection{Observational Likelihood} \label{sss:likely} 

Here we will detail how likely such a stellar system as described above is to exist, and additional considerations for its detection. We first start with an estimate of the total number of stars within 100 pc, our local interstellar neighborhood. There are $\sim$70 known stars within 5 pc\footnote{\url{http://www.recons.org/TOP100.posted.htm}}, and under the assumption that this stellar number density does not vary significantly, we estimate that there are 560,000 stars within 100 pc from our Solar System. F-type stars make up $\sim$3\% of stars in the galaxy, totaling to 16,800 stars. 
\par
Based on stellar radial velocities from GALAH DR3 \citep{2021MNRAS.506..150B,2017MNRAS.464.1259K,2021MNRAS.508.4202Z} and distances to these stars from Gaia DR2 \citep{2016A&A...595A...1G,2018A&A...616A...1G,2018A&A...616A...9L}, we estimate the percent of stars within 100 pc with radial velocities larger than $\pm$80 km s$^{-1}$ to be $\sim$1.69\%. Within 100 pc, we do not expect ISM \ion{H}{1} column densities to attenuate Ly$\alpha$ emission beyond 80 km s$^{-1}$ \citep{2022ApJ...926..129Y}. This further reduces the total number of candidate stars to 285 stars. Of these stars, we expect 24\% to be host stars to habitable exoplanets \citep{2019arXiv191206219T}. We estimate that the total number of high RV F-type stars with a habitable exoplanet within 100 pc to be 68 stars. Finally, we include the probability of such planets to be transiting. Due to their large orbital distances, the transit probability is low, $\sim$0.4\%, diminishing the number of possible candidates to zero stars within 100 parsecs. 
\par
This outcome is not ideal, however there are other cases which have not been considered. This calculation assumed the radial velocity of the ISM along a line of sight to be 0 km s$^{-1}$, however this velocity is known to vary between $\pm$ 50 km s$^{-1}$ \citep{2005ApJS..159..118W,airglow}. If our stellar radial velocity criterion is reduced to $\pm$ 30 km s$^{-1}$, there would be up to four potential candidate stars which could be observed with a fortuitous alignment of radial velocities. Additionally, there are pockets within the ISM for which column densities beyond 100 pc are not strong enough to fully attenuate Ly$\alpha$ emission beyond 80 km s$^{-1}$. Such examples are the lines of sight to $\beta$ and $\epsilon$ CmA (132 and 151 pc, respectively) which were observed by the Dual-channel Extreme Ultraviolet Continuum Experiment (DEUCE) sounding rocket \citep{2021JATIS...7a5002E}. At a distance of 150pc, there would be one potential candidate under the 80 km s$^{-1}$ constraint, and up to thirteen potential candidates under the 30 km s$^{-1}$ constraint.  
\par
We repeat the simulated observation described in Section \ref{sss:lumos} for a star located 150 pc away. The minimum exposure time to reach an SNR of 100 would be 1400 s, and coadding exposures to a total of 15 ks would result in a 3$\sigma$ detection of a planetary atmosphere within a single transit. A $\sim$6 m class space telescope equipped with an FUV spectrograph with capabilities similar to the LUMOS/G120M grating mode would theoretically be able to observe these candidate stars and their habitable exoplanets, should these star systems be identified within a myriad of stars.  

\section{Conclusions} \label{s:conclusions} 

As the characterization of a large sample of exoplanet atmospheres begins, including those of terrestrial worlds, it becomes increasingly important to characterize the SEDs of exoplanet host stars. The types of photochemical reactions, as well as exospheric properties, of exoplanet atmospheres are heavily dependant on the spectral type and activity level of its host star. Stellar FUV emission features, especially Ly$\alpha$, are the primary drivers of atmospheric photochemistry, and must be understood to contextualize the transmission spectra of exoplanet atmospheres.
\par
The SISTINE sounding rocket was designed to observe exoplanet host stars and capture their FUV radiation environments with a broad bandpass of 1010 - 1270 and 1300 - 1565 \AA. The second flight of the sounding rocket payload, SISTINE-2, successfully observed the nearby F-type star Procyon A. The flux calibrated flight spectrum was used to develop the first full SED of a mid-F main sequence star, filling a critical gap of planetary atmosphere models where an F-type host star is considered. Additionally, SISTINE-2 demonstrates several FUV technologies in preparation for an FUV spectrograph on board \textit{HWO}.
\par
We then used the newly developed SED to model the response to an Earth-like exoplanet's thermosphere and exosphere to the increased XUV radiation environment around Procyon A. We modeled various levels of stellar XUV activity, and found that the exosphere will not become hot enough to achieve an atmospheric blowoff state, and is not expected to rapidly lose its atmosphere. Transmission curves of the planetary atmosphere passing in front of its F-type host star are calculated, and are then used to simulate observations with the two current FUV spectrographs on board \textit{HST} and for an FUV spectrograph on board the future \textit{HWO}.
\par
While COS and STIS would theoretically be able to observe transits of habitable exoplanets orbiting nearby ($\sim$3.51 pc), high RV ($\pm$80 km s$^{-1}$ or larger), F-type host stars, several constraints make these detections difficult and expensive. The greatly increased effective area provided by a $\sim$6 m primary mirror makes observations of such nearby transits trivial, however we estimate very few likely candidate star systems within 100 - 150 pc for which all these conditions can be met.

\section*{Acknowledgements}
The authors thank the anonymous referee and data reviewer for their constructive questions, comments, and suggestions regarding this manuscript and its data products, the feedback has overall enhanced this work. 
\par
The authors thank current and previous members of the Colorado Ultraviolet Spectroscopy Program (CUSP) for their contributions to the design, assembly, calibration, and successful launch of the SISTINE-2 sounding rocket payload. We also thank personnel at the NASA Sounding Rocket Operations Contract (NSROC) and White Sands Missile Range (WSMR) for their support in the integration, testing, launch, and recovery of SISTINE-2. Author Cruz Aguirre thanks Allison Youngblood and Steven Cranmer for their thoughtful discussions and comments regarding this manuscript.
\par
This research was supported by NASA grant NNX16AG28G and 80NSSC20K0412.
\par
This research has made use of the NASA Exoplanet Archive, which is operated by the California Institute of Technology, under contract with the National Aeronautics and Space Administration under the Exoplanet Exploration Program.
\par
This work made use of the Third Data Release of the GALAH Survey (Buder et al. 2021). The GALAH Survey is based on data acquired through the Australian Astronomical Observatory, under programs: A/2013B/13 (The GALAH pilot survey); A/2014A/25, A/2015A/19, A2017A/18 (The GALAH survey phase 1); A2018A/18 (Open clusters with HERMES); A2019A/1 (Hierarchical star formation in Ori OB1); A2019A/15 (The GALAH survey phase 2); A/2015B/19, A/2016A/22, A/2016B/10, A/2017B/16, A/2018B/15 (The HERMES-TESS program); and A/2015A/3, A/2015B/1, A/2015B/19, A/2016A/22, A/2016B/12, A/2017A/14 (The HERMES K2-follow-up program). We acknowledge the traditional owners of the land on which the AAT stands, the Gamilaraay people, and pay our respects to elders past and present. This paper includes data that has been provided by AAO Data Central (datacentral.org.au).
\par
This work has made use of data from the European Space Agency (ESA) mission {\it Gaia} (\url{https://www.cosmos.esa.int/gaia}), processed by the {\it Gaia} Data Processing and Analysis Consortium (DPAC, \url{https://www.cosmos.esa.int/web/gaia/dpac/consortium}). Funding for the DPAC has been provided by national institutions, in particular the institutions participating in the {\it Gaia} Multilateral Agreement.

\bibliography{citations}{}
\bibliographystyle{aasjournal}

\end{document}